\pdfoutput=1
\documentclass[twocolumn,english,superscriptaddress]{revtex4-2}

\usepackage{amsmath}
\usepackage{amsfonts} 
\usepackage{babel}
\usepackage{bm}
\usepackage{color}
\usepackage{dcolumn}
\usepackage{graphics}
\usepackage{graphicx}
\usepackage[rightcaption]{sidecap}
\usepackage{helvet}
\usepackage{lineno}
\usepackage{longtable}
\usepackage{subfigure}
\usepackage{xspace}

\usepackage[colorlinks=true,linkcolor=blue,citecolor=blue,urlcolor=blue]{hyperref}

\setcitestyle{super}

\usepackage{xr}
\usepackage{xspace}

\newcommand{\average}[1]{\left\langle #1 \right\rangle}

\newcommand{\ket}[1]{\left|#1\right\rangle}

\newcommand{\units}[1]{~\text{#1}}

\newcommand{\BPCB}{(C$_5$D$_{12}$N)$_2$CuBr$_4$}
\newcommand{\BPCC}{(C$_5$D$_{12}$N)$_2$CuCl$_4$}

\newcommand{\HZ}{2.876\units{T}}
\newcommand{\MZ}{0.5}

\begin{document}


\title{Fingerprints of supersymmetric spin and charge dynamics observed by inelastic neutron scattering}

\author{Bj\"orn Wehinger}
\email[present address: European Synchrotron Radiation Facility, 38043 
Grenoble, France, ]{bjorn.wehinger@esrf.fr}
\affiliation{Department of Quantum Matter Physics, University of Geneva, 
CH-1211 Geneva, Switzerland}
\affiliation{PSI Center for Neutron and Muon Sciences, 
Paul Scherrer Institute, CH-5232 Villigen-PSI, Switzerland}

\author{Franco T. Lisandrini}
\affiliation{Physikalisches Institut, University of Bonn, 
Nussallee 12, 53115 Bonn, Germany}

\author{Noam Kestin}
\affiliation{Department of Quantum Matter Physics, University of Geneva, 
CH-1211 Geneva, Switzerland}

\author{Pierre Bouillot}
\affiliation{Department of Quantum Matter Physics, University of Geneva,
CH-1211 Geneva, Switzerland}

\author{Simon Ward}
\email[present address: Novo Nordisk A/S, Research and Early Development, 
Novo Nordisk Park 1, DK-2760  M{\aa}l\o{}v, Denmark]{}
\affiliation{PSI Center for Neutron and Muon Sciences, 
Paul Scherrer Institute, CH-5232 Villigen-PSI, Switzerland}

\author{Benedikt Thielemann}
\affiliation{PSI Center for Neutron and Muon Sciences, 
Paul Scherrer Institute, CH-5232 Villigen-PSI, Switzerland}

\author{Robert Bewley}
\affiliation{ISIS Facility, Rutherford Appleton Laboratory, Chilton, Didcot, 
Oxford OX11 0QX, United Kingdom}

\author{Martin Boehm}
\affiliation{Institut Laue-Langevin, 71 avenue des Martyrs, 38042 Grenoble, France}

\author{Daniel Biner}
\affiliation{Department of Chemistry, Biochemistry and Pharmaceutical Sciences, 
University of Bern, Freiestrasse 3, CH-3012 Bern, Switzerland}

\author{Karl W. Kr\"amer}
\affiliation{Department of Chemistry, Biochemistry and Pharmaceutical Sciences, 
University of Bern, Freiestrasse 3, CH-3012 Bern, Switzerland}

\author{Bruce Normand}
\affiliation{PSI Center for Scientific Computing, Theory and Data, 
Paul Scherrer Institute, CH-5232 Villigen-PSI, Switzerland}
\affiliation{Institute of Physics, Ecole Polytechnique F\'{e}d\'{e}rale 
de Lausanne (EPFL), CH-1015 Lausanne, Switzerland}

\author{Thierry Giamarchi}
\affiliation{Department of Quantum Matter Physics, University of Geneva, 
CH-1211 Geneva, Switzerland}

\author{Corinna Kollath}
\affiliation{Physikalisches Institut, University of Bonn, 
Nussallee 12, 53115 Bonn, Germany}

\author{Andreas M. L\"auchli}
\affiliation{PSI Center for Scientific Computing, Theory and Data, 
Paul Scherrer Institute, CH-5232 Villigen-PSI, Switzerland}
\affiliation{Institute of Physics, Ecole Polytechnique F\'{e}d\'{e}rale 
de Lausanne (EPFL), CH-1015 Lausanne, Switzerland}

\author{Christian R\"uegg}
\affiliation{Department of Quantum Matter Physics, University of Geneva, 
CH-1211 Geneva, Switzerland}
\affiliation{PSI Center for Scientific Computing, Theory and Data, 
Paul Scherrer Institute, CH-5232 Villigen-PSI, Switzerland}
\affiliation{Institute of Physics, Ecole Polytechnique F\'{e}d\'{e}rale 
de Lausanne (EPFL), CH-1015 Lausanne, Switzerland}
\affiliation{Institute for Quantum Electronics, ETH Zurich, CH-8093 
H\"onggerberg, Switzerland.}


\begin{abstract}
{\bf 
Supersymmetry is an algebraic property of a quantum Hamiltonian that, 
by giving every boson a fermionic superpartner and vice versa, may underpin 
physics beyond the Standard Model. Fractional bosonic and fermionic 
quasiparticles are familiar in condensed matter, as in the spin and charge 
excitations of the \mbox{$t$-$J$} model describing electron dynamics in 
one-dimensional materials, but this type of symmetry is almost unknown. 
However, the triplet excitations of a quantum spin ladder in an applied 
magnetic field provide a supersymmetric analogue of the \mbox{$t$-$J$} 
chain. Here we perform neutron spectroscopy on the spin-ladder compounds 
\BPCB~and \BPCC~over a range of applied fields and temperatures, and apply 
matrix-product-state methods to the ladder and equivalent chain models. From 
the momentum-resolved dynamics of a single charge-like excitation in a bath of 
fractional spins, we find essential differences in thermal broadening between 
the supersymmetric and non-supersymmetric sectors. The persistence of a 
strict zone-centre pole at all temperatures constitutes an observable 
consequence of supersymmetry that marks the beginning of supersymmetric 
studies in experimental condensed matter.}
\end{abstract}

\maketitle

Supersymmetry is a property of a quantum Hamiltonian that any bosonic 
particle has a dual fermionic superpartner and vice versa.\cite{Tong} In 
quantum field theory, it sets additional constraints on the dynamics that 
can allow analytical progress at strong coupling.\cite{Bertolini} As a 
dynamical symmetry of composite particles, it is used in nuclear physics 
to classify multiplets of heavy nuclei.\cite{iachello1980,bijker2010} Despite 
the aesthetic appeal of the possibility that it is a fundamental symmetry of 
nature, experimental evidence to date suggests that it would be severely broken
at presently accessible energy scales.\cite{Bertolini,mssm} In condensed matter, 
the concept that strong, many-body correlations can produce fractional bosonic 
and fermionic collective states is well established in low dimensions, but 
possible supersymmetries between these states are rarely invoked. Beyond some 
theoretical analysis of models constructed with explicit supersymmetry,
\cite{fendley2003,hagendorf2013} its leading applications have been in 
supersymmetric field theories relevant to critical phenomena\cite{friedan1985,
bauer2013,huijse2015} and topology.\cite{grover2014} To our knowledge even a 
simple supersymmetric Hamiltonian has yet to be engineered using ultracold atoms, 
despite extant theoretical suggestions,\cite{tomka2015,tajima2021,minar2022} but 
this has been achieved in a recent experiment using trapped ions.\cite{cai2022}

The $t$-$J$ model is a paradigm for describing charged particles interacting 
strongly with an environment of correlated quantum spins, and is thought by 
some to capture the fundamental ingredients responsible for Mott physics 
\cite{imada1998} and the enduring mysteries of cuprate superconductivity.
\cite{lee2006} Unlike the closely related Hubbard model,\cite{lieb1968} 
in one dimension the $t$-$J$ model is integrable and exactly soluble (by 
Bethe-Ansatz methods) only at one ratio of the hole-hopping and 
spin-interaction parameters, $2t/J = 1$.\cite{bares1990,bares1991,essler1992} 
At this ratio the model also has explicit supersymmetry: the two fermions 
representing the SU(2)-symmetric spin sector and the boson representing the 
U(1) charge sector are superpartners, raising the symmetry to u(1$|$2). 
Despite the many intriguing consequences of supersymmetry, the field of 
correlated quantum matter has retained a strong focus on the parameter 
regime $t \gg J$, which applies in almost all materials offering real 
and separable charge and spin degrees of freedom.

\begin{figure*}[t]
\includegraphics[width=\textwidth]{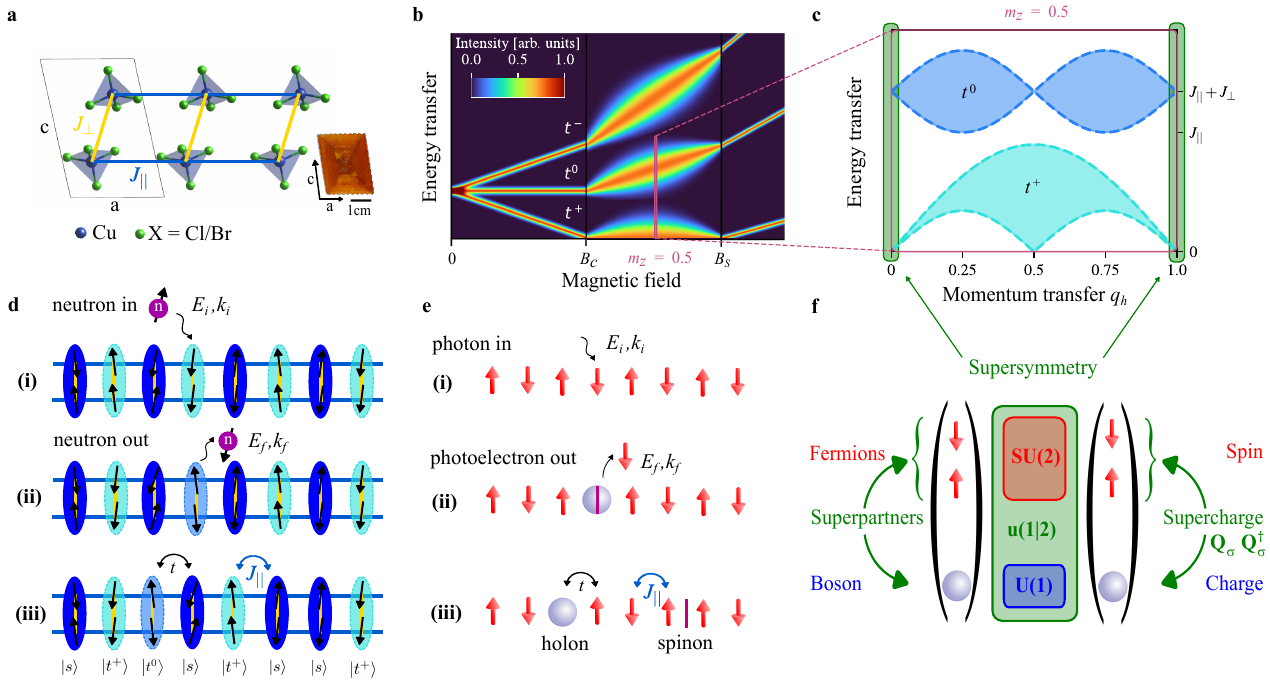}
\caption{{\bf Realization of a supersymmetric $t$-$J$ chain from a spin ladder 
in an applied magnetic field.}
{\bf a}, Schematic representation of the structure of \BPCB~(BPCB) and 
\BPCC~(BPCC) projected on the $ac$ plane, accompanied by a photograph of 
one BPCC crystal. Cu$^{2+}$ ions (blue) form the ladder and halide (Br$^-$ 
or Cl$^-$) ions (green) the superexchange paths yielding the interactions 
$J_\perp$ (yellow) and $J_\|$ (blue). The piperidinium ions are omitted 
for clarity.
{\bf b}, Evolution of triplet excitation branches in a strong-rung spin-$1/2$ 
ladder in an applied magnetic field, $B^z$. For $B_c \le B^z \le B_s$, the 
spin degrees of freedom fractionalize and the response becomes finite over a 
continuum of energies in each of the triplet branches $t^+$, $t^0$ and $t^-$
(the illustration represents an integral over all $q_h$ in panel c).  
{\bf c}, Many-body physics in the spin ladder at zero temperature. The $t^+$ 
sector gives rise to the orange two-spinon continuum. The $t^0$ sector gives 
rise to the blue spinon-holon continuum, which is described by an adapted 
$t$-$J$ model. Supersymmetry between the spinon and holon is manifest at 
wavevector $q_h = 0$ (green).
{\bf d}, Schematic representations of (i) singlet ($\ket{s} = {\textstyle 
\frac{1}{\sqrt{2}}} (\ket{\uparrow\downarrow} - \ket{\downarrow\uparrow})$) 
and triplet ($|t^+ \rangle = \ket{\uparrow\uparrow}$) states on the ladder 
rungs in the ground manifold created by the applied field; (ii) creation of a 
``middle triplon'' ($|t^0 \rangle = {\textstyle \frac{1}{\sqrt{2}}} (\ket{\uparrow
\downarrow} + \ket{\downarrow\uparrow})$) by a neutron-mediated spin excitation 
on a rung; (iii) propagation of the $|t^0 \rangle$ excitation and rearrangement 
of the $\ket{s}$-$|t^+ \rangle$ configuration. 
{\bf e}, Representation of these states in the basis of pseuospins (red 
arrows) $|\tilde{\uparrow} \rangle \equiv |{t^+} \rangle$ and $|\tilde{\downarrow}
\rangle \equiv \ket{s}$ on a chain (i), where flipping a single spin creates 
two freely mobile domain walls (spinons). Exciting a $|t^0 \rangle$ by INS 
is analogous to a photoemission process, where the injected ``hole'' is 
accompanied by one spinon (ii), both of which can propagate separately (iii). 
{\bf f}, Representation of supersymmetry as an extended symmetry in the space 
of the three site-basis states of the pure $t$-$J$ model at $J = 2t$: the 
additional duality between the fermionic spins (with SU(2) spin symmetry) 
and the bosonic hole state (with U(1) charge symmetry), which makes them 
superpartners, is encoded in the supercharge operators $Q_\sigma$ and 
$Q_\sigma^{\dag}$.}
\label{fig:intro}
\end{figure*}

In experiment, the physics of a single charge in a magnetic environment has 
been probed by photoemission spectroscopy, and signatures of spin-charge 
separation have been reported in 1D metals,\cite{segovia1999,claessen2002} 
insulators\cite{kim1996,kim1997,kim2006} and cold-atom systems,
\cite{vijayan2020} as has spin-orbital separation in a 1D insulator.
\cite{schlappa2012} In this context, low-dimensional quantum magnets offer 
a wide variety of exotic physical properties, generically low (meV) energy 
scales accessible to laboratory magnetic fields and high-intensity, 
high-resolution measurements of the full spectral function by modern 
inelastic neutron scattering (INS) spectrometers. Magnetic-field control 
of the quasiparticle density in the two-leg quantum spin ladder, represented 
in Figs.~\ref{fig:intro}a-c, allowed the quantitative testing of phenomena 
such as field-induced Bose-Einstein condensation and the formation of the 
spin Tomonaga-Luttinger liquid (TLL), which exhibits the fractionalization of 
magnon excitations into deconfined spinons.\cite{klanjsek2008,thielemann2009a,
schmidiger2013,ward2017} It has been pointed out that, in parallel, the 
two-leg ladder in a field also provides a rather faithful realization of 
a single hole in the $t$-$J$ model, with one excitation in the middle 
Zeeman-split triplet branch ($t^0$) playing the role of this 
hole,\cite{bouillot2011} as represented in Figs.~\ref{fig:intro}d-e. 
Here we point out in that, in addition, this ladder-based \mbox{$t$-$J$} 
model is supersymmetric. 

In this Article, we investigate the consequences of supersymmetry in the 
ladder-derived \mbox{$t$-$J$} model, and in the process take the first steps 
in studying supersymmetry as an experimental science. The basic tenets 
of supersymmetry as an extended symmetry of the \mbox{$t$-$J$} basis are very 
simply stated, but to understand their manifestations in a condensed-matter 
system we proceed from ladder materials through INS experiments and numerical 
many-body calculations. Specifically, we perform high-resolution INS 
measurements of two quantum spin-ladder materials, using the applied 
magnetic field and the temperature as control parameters. We focus on the 
spectral functions of the middle ($t^0$) triplet branch in the two parity 
sectors of the ladder to highlight the presence or absence of supersymmetry 
between the charge and spin fractions of the electron in the equivalent 
\mbox{$t$-$J$} chain. For a quantitative analysis of our results, we perform 
zero- and finite-temperature matrix-product-states (MPS) calculations both for 
the ladder and for selected \mbox{$t$-$J$} models. We show how the presence of 
an exact pole in the ladder spectrum has consequences in the \mbox{$t$-$J$} 
model that allow us to demonstrate observable fingerprints of supersymmetry 
in condensed matter.  

\begin{figure}[t]
\includegraphics[width=0.98\columnwidth]{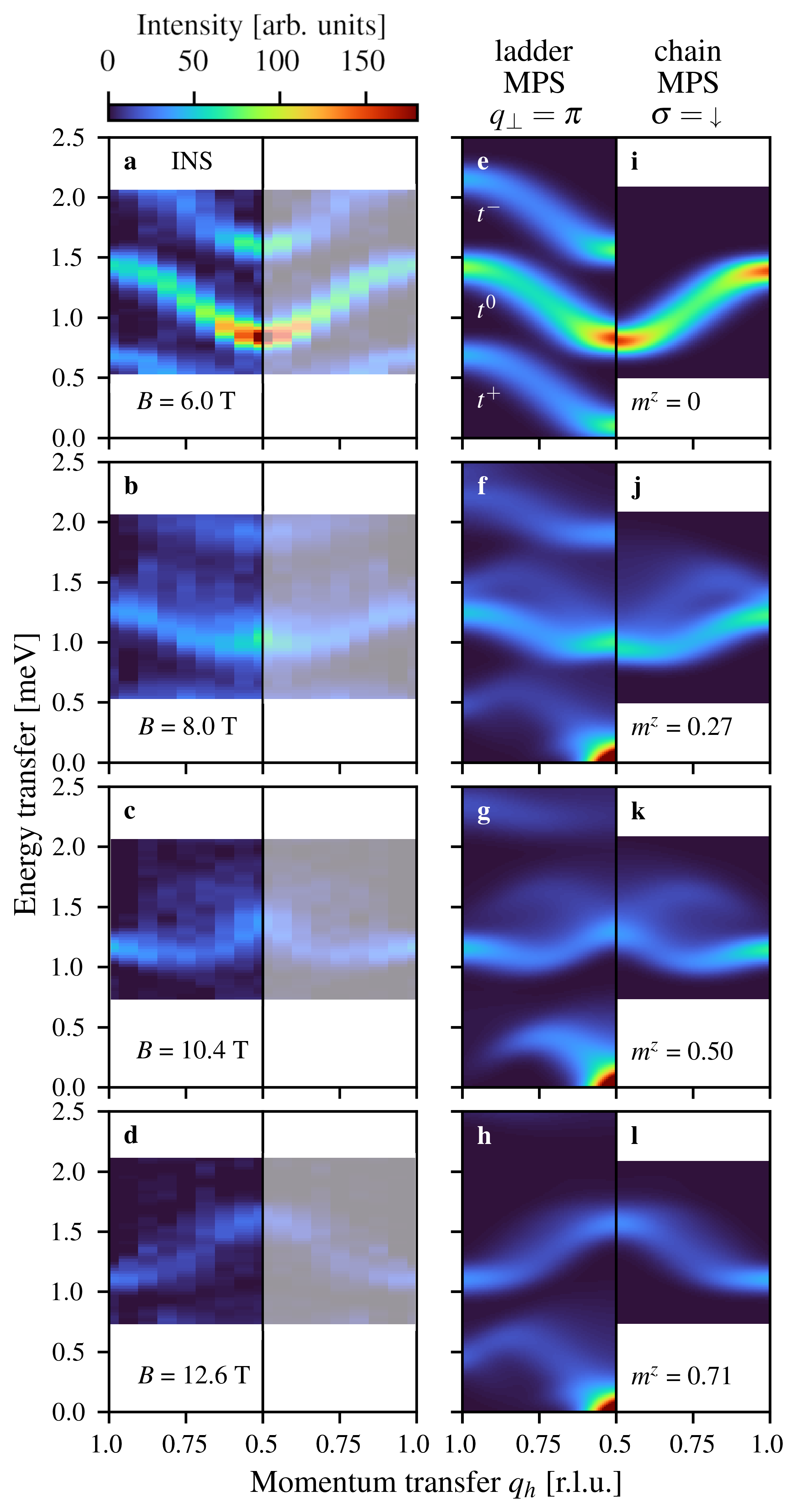}
\caption{{\bf Field-induced evolution of the magnetic excitation spectrum in 
BPCB.} {\bf a}-{\bf d}, Measured neutron scattering intensities in sector 
$q_\perp = \pi$ at a temperature of 50 mK and applied magnetic fields of 
6.0, 8.0, 10.4 and 12.6~T, where the ladder magnetizations are respectively 
$m^z = 0$, 0.27, 0.50 and 0.71. Spectra in the faded region were generated 
by symmetry.
{\bf e}-{\bf h}, Corresponding spectral functions obtained from MPS 
calculations performed for the magnetized spin ladder at zero temperature
(l-MPS).
{\bf i}-{\bf l}, Zero-temperature spectral functions of the $t^0$ sector 
calculated by MPS using the equivalent model of one hole in a modified 
$t$-$J$ chain (c-MPS).}
\label{fig:hspectra}
\end{figure}

\bigskip
\noindent
{\bf{Supersymmetry in a spin ladder}}

\vspace{4pt}

\begin{figure*}[t]
\includegraphics[width=\textwidth]{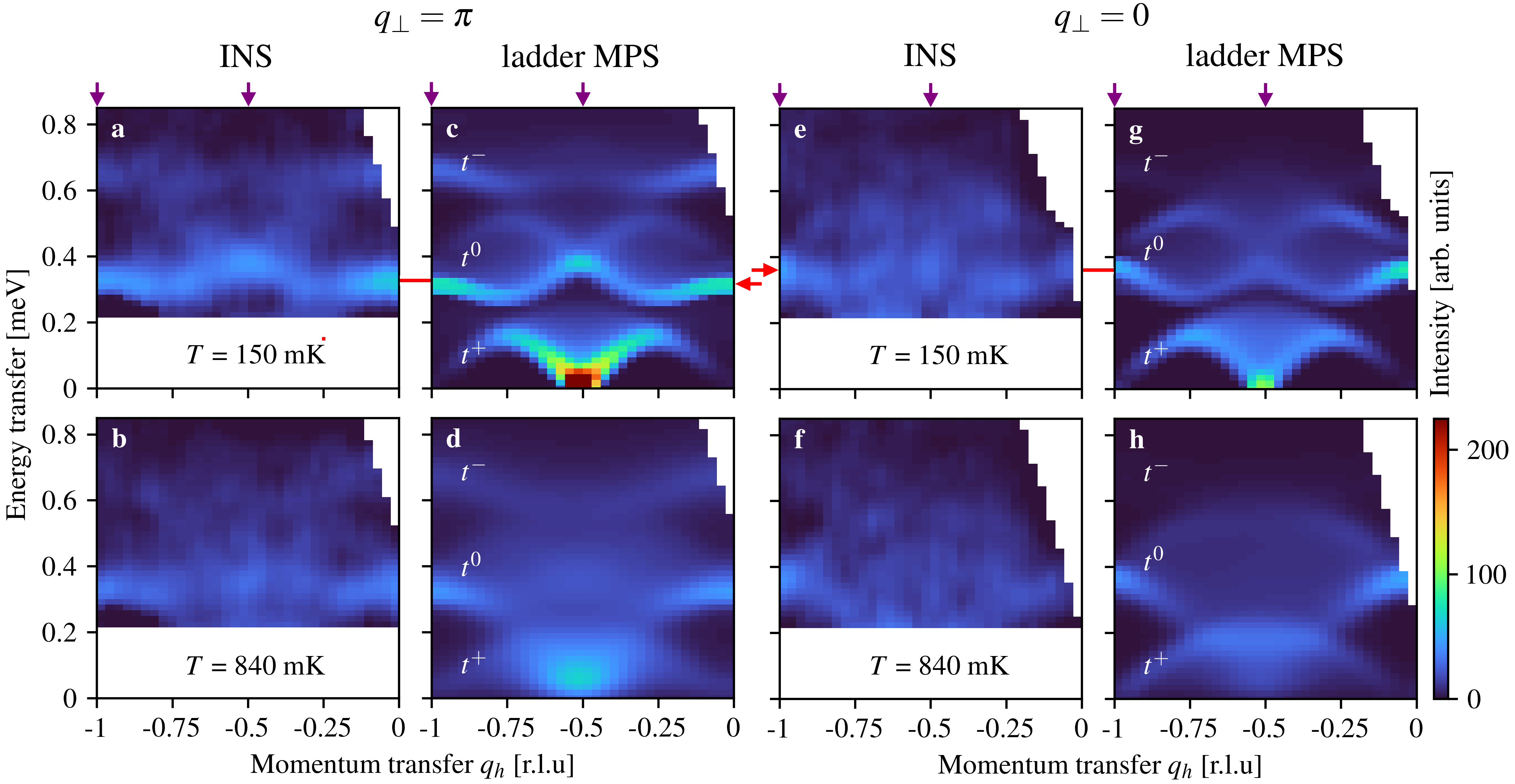}
\caption{{\bf Thermal evolution of the magnetic excitation spectrum in BPCC 
at $m^z = 0.5$ ($B^z = \HZ$)}. {\bf a}-{\bf b}, Measured neutron scattering 
intensities in sector $q_\perp = \pi$ at temperatures of 150 and 840 mK. 
{\bf c}-{\bf d}, Corresponding ladder MPS calculations at 0 and 840 mK. 
{\bf e}-{\bf f}, Measured intensities in sector $q_\perp = 0$ at temperatures 
of 150 and 840 mK. {\bf g}-{\bf h}, Corresponding ladder MPS calculations. 
Vertical arrows mark $q_h$ cuts analysed in detail and horizontal arrows 
draw attention to the differences between $t^0$ continua in the two $q_\perp$ 
sectors.}
\label{fig:spectra}
\end{figure*}

To realize supersymmetric physics we consider the deuterated compounds 
bis-piperidinium copper(II) bromide (BPCB) and chloride (BPCC), which form 
excellent two-leg quantum spin ladders. The Cu$^{2+}$ ions carry localized 
$S = 1/2$ moments and through double halide bridges form spin dimers 
constituting the ladder rungs (yellow lines in Fig.~\ref{fig:intro}a), while 
single halide bridges form ladder legs (blue lines) along the crystalline 
$a$ axis.\cite{ward2013} Neighbouring ladders are separated by the large, 
organic piperidinium ions, (C$_5$D$_{12}$N)$^+$. Crystal growth and structure 
are described in the Methods section and in Sec.~S1 of the Supplementary 
Information (SI).\cite{SI} The microscopic Hamiltonian describing the compound 
in an applied magnetic field is 
\begin{eqnarray}
\label{eq:hamiltonian_ladder}
\! \! \! \! \! \! H & \! = \! & J_\perp \!\! \sum_{i} \! \vec{S}_{i,1} \! 
\cdot \! \vec{S}_{i,2} + J_\| \! \sum_{i,\eta} \vec{S}_{i,\eta} \! \cdot \! 
\vec{S}_{i+1,\eta} - b^z \! \sum_{i,\eta,} \! S^z_{i,\eta},
\end{eqnarray}
in which the spin operator $\vec{S}_{i,\eta}$ acts at site $i$ of leg $\eta 
\in \{1,2\}$ in the ladder and $b^z = g \mu_B B^z$ with $\mu_B = 0.672\units
{K/T}$ the Bohr magneton, $g$ the gyromagnetic factor and $B^z$ the applied 
magnetic field. The antiferromagnetic Heisenberg interaction parameters 
determined by INS for BPCB are $J_\perp = 12.6\units{K}$ on the ladder rung 
and $J_\parallel = 3.55\units{K}$ on the leg, with $g = 2.28$ for a field 
parallel to the $b$ axis,\cite{ward2013} while for BPCC $J_\perp = 
3.42\units{K}$, $J_\parallel = 1.34\units{K}$ and $g = 2.26$.\cite{ward2017} 
Thus both materials are strong-rung ladders with a similar ratio $J_\perp/ 
J_\parallel$, which is well adapted for separating the triplet excitation 
sectors by energy in an applied field. 

The energy scales of $J_\perp$ and $J_\parallel$ in both materials are well 
suited for high-precision INS measurements, including in fields up to 
saturation ($B_s = 13.79$ T in BPCB and 4.07 T in BPCC). The lower energy 
scale in BPCC facilitates working at finite temperatures without incurring 
background comparison problems. The interladder coupling $J' \lesssim 0.002 
J_\perp$ in both cases\cite{thielemann2009,ward2017} is sufficiently small 
that we focus on the properties of isolated ladders 
[Eq.~\eqref{eq:hamiltonian_ladder}]. INS experiments were performed on the 
triple-axis spectrometer THALES at the Institut Laue Langevin (ILL) and on 
the time-of-flight spectrometer LET\cite{bewley2011} at the ISIS pulsed 
neutron source (Methods section). The former were prioritized to probe 
ladder physics at different applied fields and the latter to probe the 
spectral response with optimal energy and momentum resolution. The elegant 
parity selectivity of the ladder geometry allows a complete separation 
of singlet-triplet excitation processes, which appear in the antisymmetric 
rung momentum sector ($q_\perp = \pi$), and triplet-triplet processes, which 
appear in the symmetric sector ($q_\perp = 0$);\cite{bouillot2011,ward2017} 
a detailed discussion is provided in Supplementary Note 2.\cite{SI} 

As noted above, the physics of a single doped hole in a $t$-$J$ chain 
appears in the two-leg spin ladder when describing the response of the $t^0$ 
sector, depicted in blue in Fig.~\ref{fig:intro}c.\cite{bouillot2011} As 
Figs.~\ref{fig:intro}d-e represent schematically, one $t^0$ excitation 
appears as a hole $\ket{0} \equiv |t^0 \rangle$ in the background of pseudospins 
$|\tilde{\uparrow} \rangle$ and $|\tilde{\downarrow} \rangle$ representing the 
$\ket{s}$-$|t^+ \rangle$ ground-state condensate at fields $B_c \le B \le B_s$. 
For the Hamiltonian of Eq.~\eqref{eq:hamiltonian_ladder}, this $t$-$J$-chain 
model takes the form 
\begin{eqnarray}
\label{eq:1dtj}
H_{tJl} & = & t \sum_i (c^{\dagger}_{i\tilde{\uparrow}} c_{i+1\tilde{\uparrow}}
 + c^{\dagger}_{i\tilde{\downarrow}} c_{i+1\tilde{\downarrow}} + \text{H.c.}) \\
&& + J_{\rm I} \sum_i ( n_{i h} n_{i+1} \! + \! n_i n_{i+1 h})
 + \mu \sum_i \! n_{i h} + H_{\rm XXZ}, \nonumber
\end{eqnarray}
where 
$t = {\textstyle \frac12} (J_{\|}$, $J_{\rm I} = - {\textstyle \frac{1}{4}} 
J_{\|}$ and $\mu = {\textstyle \frac12} (b^z + J_{\perp})$. $c^{(\dagger)}
_{i\sigma}$ annihilates (creates) a fermion at site $i$ with pseudospin 
$\sigma = \tilde{\uparrow},\tilde{\downarrow}$, $n_{i}$ is the fermion 
number operator and $n_{i h}$ is the hole number operator. In the absence 
of a hole excitation, the pseudospin-1/2 system has effective Hamiltonian
\begin{equation}
\label{eq:xxz}
H_{\rm XXZ} = \! \sum_i \! \big[ J_{\|} (S^x_i S^x_{i+1} \! + \! S^y_i 
S^y_{i+1}) \! + \! {\textstyle \frac12} J_\| S^z_i S^z_{i+1} \big]
 - h \! \sum_i \! S^z_i,
\end{equation}
with $h = b^z - J_{\perp} - {\textstyle \frac12} J_\|$. The $t^0$ sector is then 
described by a specific form of $t$-$J$ chain with an XXZ spin anisotropy and an 
interaction term, $J_{\rm I}$, between a hole and a $\tilde{\uparrow}$ spin. A 
more systematic and general derivation of this model is presented in Supplementary 
Note 3.\cite{SI} We stress that Hamiltonians \eqref{eq:hamiltonian_ladder} and 
\eqref{eq:1dtj} provide almost identical descriptions of the $t^0$ sector, differing 
only by minor corrections at ${\cal{O}} [(J_\|/J_\perp)^2]$: we use the model of 
Eq.~\eqref{eq:hamiltonian_ladder} in all of the ladder calculations to follow and 
the model of Eq.~\eqref{eq:1dtj} in all of the $t$-$J$-chain calculations, except 
when explicitly stated otherwise.

Returning now to the topic of supersymmetry, despite the mystique of its
predictions beyond the Standard Model, supersymmetry {\it per se} is simply 
an extended symmetry group that relates the operators in a Hamiltonian.
\cite{Tong} The supersymmetric parameter ratio of the \mbox{$t$-$J$} chain 
studied in Refs.~\cite{bares1990,bares1991,essler1992} is obtained rather 
straightforwardly from the fact that the ladder model has only one energy 
scale for inter-rung processes, namely $J_{\|}$. Technically, the 
\mbox{$t$-$J$} chain with a Heisenberg spin sector \cite{bares1990,
bares1991,essler1992} has two supersymmetries, between the hole and 
the up-spin and between the hole and the down-spin, as depicted in 
Fig.~\ref{fig:intro}f. With two fermions (F) and one boson (B), the 
system is equivalent to an FFB permutation model. \cite{sutherland1975} 
Quite generally, the FFB-type model with three fully symmetric species has 
9 symmetry operators that make up the Lie superalgebra u(1$|$2), which we 
specify in full in Supplementary Note 3.\cite{SI} While one of these operators 
is the identity, one is U(1) charge conservation and three are the SU(2) 
spin symmetry, the four additional operations, written as the ``supercharge'' 
operators $Q_\sigma$ and $Q_\sigma^\dag$, connect the bosonic hole to the 
fermionic spins and hence make these superpartners. In the ladder-derived 
\mbox{$t$-$J$} model of Eq.~\eqref{eq:1dtj}, the two additional interaction 
terms combine to remove the second supersymmetry, and we will show in theory, 
numerics and experiment how the system retains one exact supersymmetry with 
operators $Q_{\tilde{\uparrow}}$ and $Q_{\tilde{\uparrow}}^\dag$ (Lie 
superalgebra u(1$|$1)), as depicted in Supplementary Fig.~3.\cite{SI}

\vspace{4pt}

\bigskip
\noindent
{\bf{Spin ladder in a magnetic field}}

\vspace{4pt}
To establish the baseline for our investigation, we look first at 
Fig.~\ref{fig:hspectra}a, where a field of 6~T applied to BPCB causes Zeeman 
splitting of the three gapped magnon modes, which we label from bottom to top 
as $t^+$, $t^0$ and $t^-$. When the $t^+$ gap closes, the excitations 
become gapless spinons\cite{thielemann2009a} and the low-energy sector of the 
half-magnetized ladder shows the well known des Cloizeaux-Pearson-Faddeev 
(dCPF)\cite{dcp} continuum spectrum (turquoise shading in Fig.~\ref{fig:intro}c). 
The field functions as an effective chemical potential for the spinons (depicted
in Supplementary Fig.~4 and discussed in Supplementary Note 4\cite{SI}) and 
is reflected directly in the incommensurability of the spinon continuum 
(Figs.~\ref{fig:hspectra}f,h). A $t^0$ excitation constitutes a third type of 
particle dressed by a spinon, exactly analogous to a single hole in a spin chain 
(Figs.~\ref{fig:hspectra}j-l), and the supersymmetric relationship between 
the hole and the spinon fractions of the $t^0$ branch is the object of our 
current investigation. Figure \ref{fig:hspectra} shows how raising the 
field through the gapless regime changes the shape of the $t^0$ continuum 
(Figs.~\ref{fig:hspectra}b-d): while the spectral weight remains concentrated 
between 1.0 and 1.7 meV, the minima and maxima move systematically across the 
Brillouin zone as the field changes the spin polarization (and hence the Fermi 
wavevector of the effective spinons). 

To analyse our measured spectra, we compare our BPCB results directly with 
the time-dependent MPS calculations\cite{white2004,daley2004,schollwock2011} 
of Ref.~\cite{bouillot2011}, and have performed analogous calculations for 
the parameters of BPCC. A summary is provided in the Methods section and 
further details of the results in Supplementary Note 5.\cite{SI} Figures 
\ref{fig:hspectra}e-h show how the emergence of deconfined spinons at the 
lower critical field ($B_{c} = 6.6$~T in BPCB) causes all three magnon 
branches to decompose into continua whose intensity structures evolve 
systematically with the field. To interpret the features of the $t^0$ 
continuum, we have performed additional MPS calculations for the dynamics 
of a single hole in different variants of the $t$-$J$ model (Supplementary 
Note 6\cite{SI}). Figures \ref{fig:hspectra}e-l make clear the 
quantitative equivalence of the two models for the $t^0$ branch.

For a comprehensive experimental analysis of the $t^0$ continuum, and 
hence of ladder-derived $t$-$J$ physics, we combined the capabilities of 
the LET spectrometer with the lower energy scales of BPCC to extend our 
investigation in two ways. Working exclusively at half magnetization 
($m^z = 0.5$, $B^z = 2.876$ T, analogous to Figs.~\ref{fig:hspectra}c, 
\ref{fig:hspectra}g and \ref{fig:hspectra}k), we measured the symmetric 
($q_\perp = 0$) sector of the ladder with the same degree of accuracy as 
the antisymmetric ($q_\perp = \pi$) sector studied in Fig.~\ref{fig:hspectra}. 
Our low-temperature results (top row of Fig.~\ref{fig:spectra}) show two $t^0$ 
continua with clearly different intensity structures and discernibly different 
energies, but sharing the common features that most intensity is concentrated 
along their lower edges and their maximal spectral weight occurs at $q_h \equiv 
q_\parallel /2 \pi = 0$ and $- 0.5$. The second extension is to repeat our 
measurements at finite temperatures (on the order of $J_\|$, bottom row of 
Fig.~\ref{fig:spectra}), and we will show why both extensions are key to 
revealing the effects of supersymmetry. 

In the spin ladder it is known from MPS calculations\cite{bouillot2011} that 
the two $t^0$ continua differ, although an experimental measurement 
(Fig.~\ref{fig:spectra}) was not previously available. These differences 
can be regarded as a consequence of interaction effects beyond the simple 
picture of convolving a $t^0$ and a spinon (Fig.~\ref{fig:intro}c); spinon 
interactions in the $t^+$ branches\cite{keselman2020} are discussed in 
Supplementary Note 4.\cite{SI} Before interpreting their effects in the two 
$t^0$ continua, we stress a little known but fundamental property of any 
Heisenberg model in an applied field, that the spectral function at wavevector 
${\bf q} = {\bf 0}$ has an exact pole at the Zeeman energy.\cite{oshikawa1997,
keselman2020} In our BPCC measurements and MPS spectra, the continuum 
response in the $q_\perp = 0$ sector sharpens dramatically at $q_h = 0$ 
because of this pole.  

To analyse holon and spinon interaction effects in the two $t^0$ continua, we 
turn to the mapping of the ladder to the properties of one hole in the $t$-$J$ 
chain specified in Eq.~\eqref{eq:1dtj}.\cite{bouillot2011} 
The key properties of the mapping are (i) the spin chain has 
XXZ interactions with $J_z = J_\|/2$; (ii) the hole hopping is given by $t = 
J_\|/2$; (iii) an interaction term, $J_{\rm I} = J_\|/4$, appears only 
between the hole and an up-spin. Property (i) gives the XXZ chain spinon 
excitations and leads to the ``noninteracting'' interpretation of the 
measured $t^0$ spectral function (Fig.~\ref{fig:spectra}) as a convolution 
of the elementary $t^0$ branch (the cosine observed at $B = 0$) with a single 
spinon of Fermi wavevector $q_\parallel = \pm \pi/2$ (blue shading in 
Fig.~\ref{fig:intro}c).\cite{sorella1998,bouillot2011} Property (iii) 
distinguishes between hole ($t^0$) states of $|s \rangle$ ($\tilde{\downarrow}$) 
or $|t^+ \rangle$ ($\tilde{\uparrow}$) origin in the ladder, causing the two 
spectral functions to develop the differences we observe between the symmetric and 
antisymmetric sectors (shown explicitly by MPS in Supplementary Fig.~5\cite{SI}). 
Property (ii) establishes the two supersymmetries of the pure $t$-$J$-chain 
model, and in Supplementary Fig.~S7\cite{SI} we illustrate how all three 
properties combine to preserve a single supersymmetry of the ladder-derived 
$t$-$J$ model, which is the duality between the $\tilde{\uparrow}$ fermion 
and the bosonic holon. 

\begin{figure}
\includegraphics[width=\columnwidth]{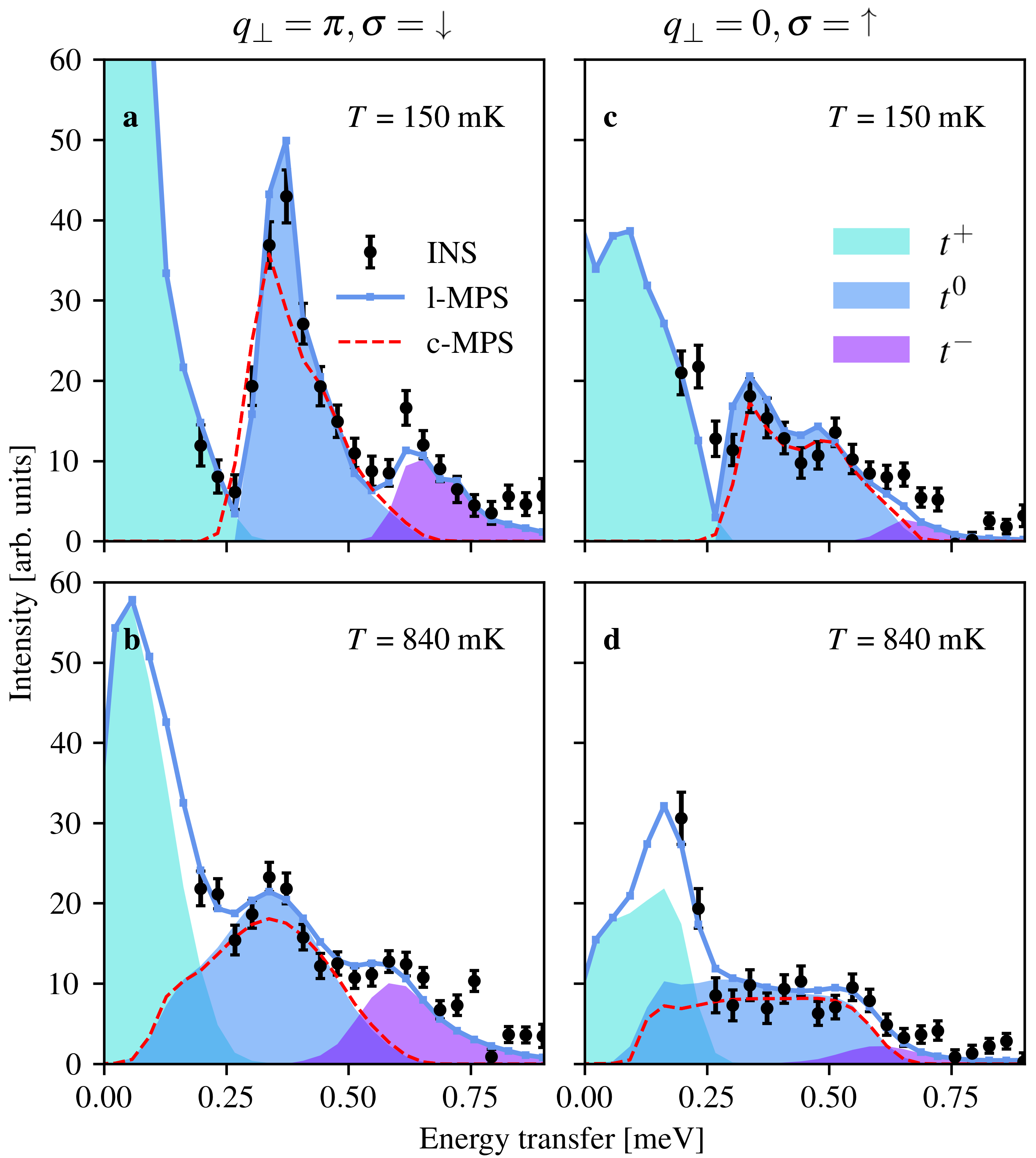}
\caption[]{{\bf Many-body thermal broadening.} Background-subtracted scattered 
intensities measured for BPCC at $m^z = \MZ$ and at low (a,c) and intermediate 
(b,d) temperatures, integrated over momentum transfers $q_h \in [-0.625,
-0.375]$ and shown for sectors $q_\perp = \pi$ (a,b) and $q_\perp = 0$ (c,d). 
In both $q_\perp$ sectors one observes a dramatic increase in the spectral 
weight of the $t^0$ continuum at energies below the $T = 0$ edge. The solid 
blue line shows the ladder spectrum (l-MPS) and the dashed red line the 
spectral function of a single hole in a $t$-$J$ chain (c-MPS), both computed 
by zero-temperature MPS in the upper row and by finite-temperature MPS in the 
lower row, and with the same momentum integration, data binning and equivalent 
resolution as in experiment.}
\label{fig:slices4} 
\end{figure}

\bigskip
\noindent
{\bf{Observable consequences of supersymmetry}}

\vspace{4pt}

To search for experimental observables, a first essential statement is that 
supersymmetry is a global symmetry: the supersymmetric operators act on the 
entire system, and summing over all lattice sites makes them relevant at 
wavevector ${\bf q} = {\bf 0}$. Second, the action of these operators in 
interchanging bosons and fermions allows exact statements about the energies 
in sectors of different particle number. Explicitly, if $|\psi(N_e)\rangle$ 
is any eigenstate of $N_e$ $\tilde{\uparrow}$ particles with energy $E$, then 
$Q_{\tilde{\uparrow}}^{(\dag)} |\psi(N_e)\rangle$ is an exact eigenstate of $N_e
 \pm 1$ particles with energy $E \pm 2t$ (assuming $\mu = 0$).\cite{bares1991} 
While other authors have considered sectors of arbitrary $N_e$,\cite{bares1991,
saiga1999} in the ladder-derived $t$-$J$ chain this establishes an exact  
correspondence between the zero- and one-hole sectors. Because the 
single-hole or -electron dynamical spectral function involves the matrix elements 
of $c^{(\dag)}_{q\sigma}$ (which reduce to $Q^{(\dag)}_{\sigma}$ at ${\bf 
q} = {\bf 0}$) between the two sectors, the fact that all states in each sector 
have the same energetic separation from their partners guarantees that the 
spectral function has an exact $\delta$-function at zero momentum and at an 
energy close to $\pm 2t$. Because the thermal states of both sectors still 
have the same supersymmetric relationship, the $\delta$-function form persists 
at all temperatures (up to the limit of the ladder-chain mapping, which is set 
by $J_\perp$). The weight of this $\delta$-function may depend on the 
temperature or magnetic field, but its nature and location in energy are 
fixed by the supersymmetry. Supersymmetry thus appears as another symmetry 
protecting the response, in this case from thermal broadening.

We stress that, when considering the electronic response as a convolution of 
spin and hole degrees of freedom in a $t$-$J$ chain, the shape of the support 
is not obvious from kinematics and hence the supersymmetric $\delta$-function 
is most intriguing. However, the fact that the model is derived from a 
Heisenberg Hamiltonian in a uniform magnetic field guarantees a single pole 
in the response function; this pole is inherited by the chain model we study, 
although for the most general Heisenberg ladder model (Supplementary Note 
3C\cite{SI}) it can be found at different combinations of $t_\sigma/J$, 
the XXZ anisotropy and the strength of the interaction term. 

\begin{figure}
\includegraphics[width=\columnwidth]{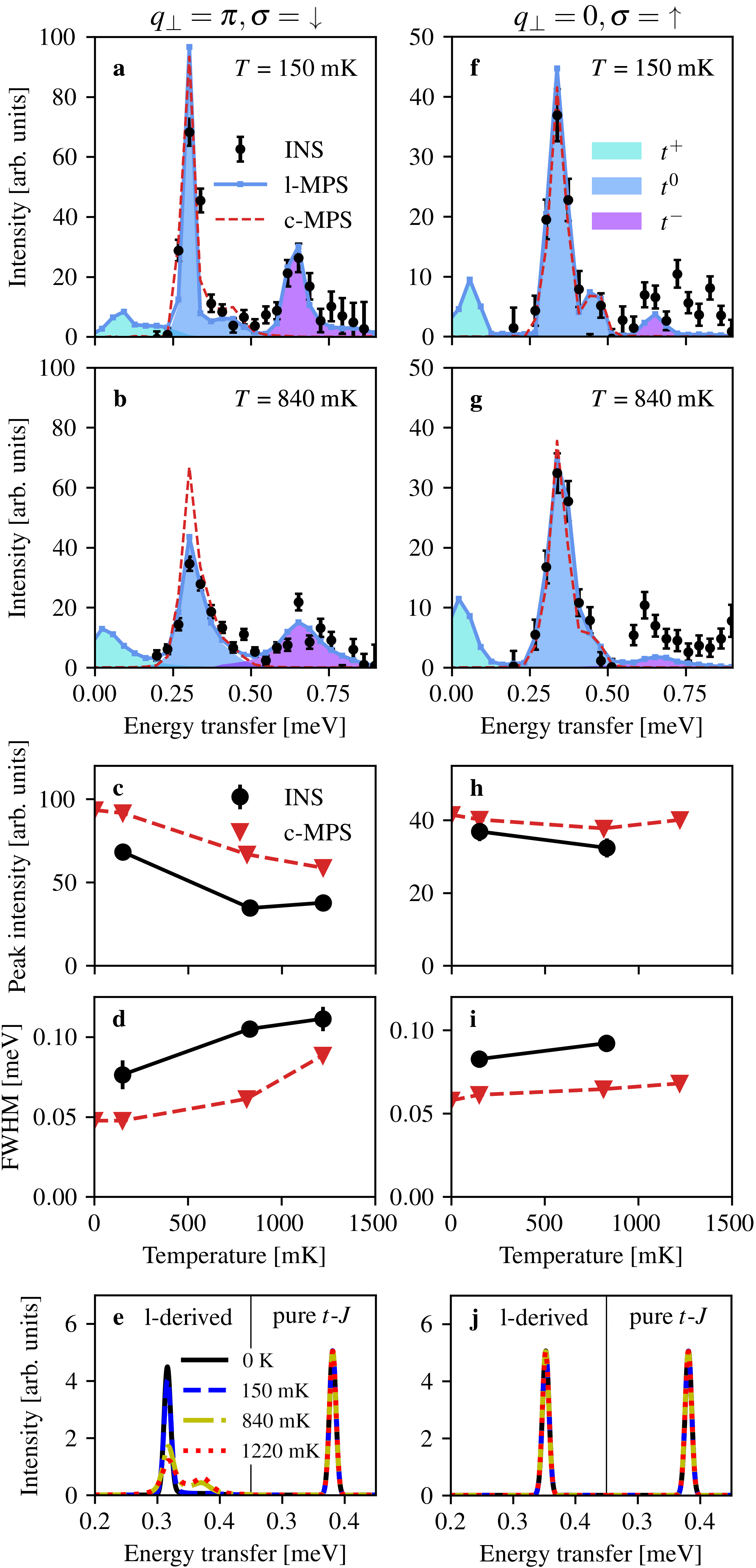}
\caption[]{{\bf Supersymmetric protection from thermal broadening.} 
{\bf a-b},{\bf f-g}, Background-subtracted scattered intensities measured for 
BPCC at $m^z = \MZ$ at low (a,f) and intermediate (b,g) temperatures, 
integrated over $q_h \in [-1.125,$ $-0.875]$ and shown for sectors $q_\perp = 
\pi$ (a,b) and $q_\perp = 0$ (f,g). The l-MPS and c-MPS lines are as in 
Fig.~\ref{fig:slices4}.
{\bf c-d},{\bf h-i}, Heights and widths of the $t^0$ spectral peaks shown 
as functions of temperature and compared with c-MPS calculations. Without 
supersymmetry, increasing $T$ causes a reduced intensity (c) and broader 
peak (d), whereas in the supersymmetric sector there is no thermal 
evolution within the experimental and numerical resolution. 
{\bf e},{\bf j}, Spectral functions of the pure and ladder-derived $t$-$J$ 
chains computed at $q_h = 0$ (i.e.~with no momentum integration) in both 
parity ($q_\perp$) sectors at $T = 0$ and for the three experimental 
temperatures. } 
\label{fig:slices5} 
\end{figure}

Finding fingerprints of supersymmetry (appearing around $q = 0$) thus 
requires a comparison with the conventional kinematics of thermally induced 
many-body phenomena (appearing at generic $q$ values). In experiment, strong 
thermal effects on conventional spin-charge separation have been reported in 
the response of a 1D conductor.\cite{claessen2002} Numerically, computing the 
spectrum of a 1D system at finite temperatures poses a challenge for MPS 
methods in controlling the spread of entanglement. We have implemented 
state-of-the-art MPS techniques\cite{schollwock2011,barthel2013,kestin2019} 
based on density-matrix purification\cite{verstraete2004,zwolak2004} to 
obtain accurate numerical results for both the ladder and chain models at 
the experimental temperatures. Because our MPS spectra offer additional 
insight into many-body thermal phenomena that are obscured in the INS spectra, 
particularly below 0.15 meV, we discuss the $t^+$ continua in Supplementary 
Note 4 and the $t^0$ (\mbox{$t$-$J$}) continua in more detail in Supplementary 
Notes 5 and 6.\cite{SI}

The most obvious thermal effect on the $t^0$ continua is the rapid 
loss of the narrow, intense low-$T$ features in both parity sectors 
[Figs.~\ref{fig:spectra}a,c,e,g] as all spectra undergo a dramatic broadening 
towards lower energies around $q_h = - 0.5$ [Figs.~\ref{fig:spectra}b,d,f,h]. 
Some weight is also shifted upwards, creating wide distributions that become 
nearly uniform over their entire energy range. For a quantitative analysis, 
we discretize $q_h$ into three regimes by integrating over regions of width 
$\Delta q_h = 0.25$ centred at $q_h = -1$, $q_h = - 0.75$ and $q_h = - 0.5$; 
the most striking example of this many-body thermal broadening occurs around 
$q_h = - 0.5$ and $q_\perp = \pi$ [Figs.~\ref{fig:slices4}a,b], where the 
high-energy half of the $t^0$ continuum changes rather little, whereas the 
low-energy half shows a striking shift of weight from the maximum, essentially 
filling in the formerly empty energy window from 0.15 to $0.3\units{meV}$. 
The changes in the $q_\perp = 0$ sector [Figs.~\ref{fig:slices4}c,d] are 
qualitatively the same, as in fact is the physics of the $t^-$ continuum. 

Although far from the conventional thermal broadening of an integer-spin mode, 
this physics is generic for fractionalized excitations. At finite temperatures,
the leading higher-order terms in the $t^0$ spectrum involve three spinons. The
spectral response is therefore broadened vertically by all possible energies of
the additional spinon pair,\cite{saiga1999} as depicted alongside MPS data in 
Supplementary Fig.~5,\cite{SI} and the most obvious consequence at finite $T$ 
is to occupy additional initial states, which accounts directly for the strong 
spectral weight in the range $\omega = 0.15$-0.3 meV at $q_h = - 0.5$ in 
Figs.~\ref{fig:slices4}b,d. A more detailed discussion is deferred to 
Supplementary Note 5, where Supplementary Fig.~6 provides the experimental 
data for thermal broadening in the mid-zone regime (i.e.~around $q_h
 = - 0.75$).\cite{SI} 

At $q_h = 0$ (or equivalently $-1$, Fig.~\ref{fig:slices5}) we anticipate 
a special situation, whose confirmation is our central result. Here the 
supersymmetry should protect a $\delta$-function response at all temperatures 
for $q_\perp = 0$, but not for $q_\perp = \pi$. Observing this effect is 
complicated by the fact that the finite-$T$ $q_\perp = \pi$ peak 
(Fig.~\ref{fig:slices5}b) is already rather narrow compared to $q_h \ne 0$ 
(Fig.~\ref{fig:slices4}c). This is a consequence of kinematic effects also 
visible in Fig.~\ref{fig:spectra}, and is not a hallmark of any close proximity 
to supersymmetry (we recall that $J_{\rm I}$ is on the order of $J_\|$). 
Nevertheless, this peak becomes half as tall and 50\% broader from 150 to 
840 mK (Figs.~\ref{fig:slices5}a-d). By contrast, the peak at $q_\perp = 0$ 
in Figs.~\ref{fig:slices5}f-g changes by less than 15\% in both measures 
(Figs.~\ref{fig:slices5}h-i). 

To establish a benchmark for the properties of the measured peaks, given 
the finite momentum integration and instrumental resolution, we use our 
$t$-$J$-chain calculations to obtain the dashed red lines in 
Figs.~\ref{fig:slices5}a-d and \ref{fig:slices5}f-i. We remark here that 
our finite-$T$ MPS augmented by linear prediction of the time evolution 
connects smoothly from the spectral peak at 150 mK to $T = 0$. We stress 
again that our results consist of only one high-temperature dataset ($T = 
840$ mK, Fig.~\ref{fig:slices5}g) in the supersymmetric $q_\perp = 0$ sector, 
with a rather low peak intensity and a broad integration window contributing 
an extra feature around 0.44 meV. Nevertheless, the contrasting shapes of the 
primary peaks in Figs.~\ref{fig:slices5}b,g and the agreement with MPS 
modelling provide quite striking evidence for the influence of supersymmetry. 

In Figs.~\ref{fig:slices5}e,j we use our finite-$T$ MPS calculations at their 
highest resolution (approximately 0.025 r.l.u.~in momentum and 0.012 meV 
in energy) to compare the supersymmetric spectral function of the ladder-derived 
\mbox{$t$-$J$} chain with that of the pure (Heisenberg) $t$-$J$ chain of 
Refs.~\cite{bares1990,bares1991,essler1992} In the Heisenberg $t$-$J$ model, 
the peak shapes at $q_h = 0$ for both the up- and down-spin response functions are 
supersymmetry-protected against thermal evolution, showing no change in height 
or width. In the ladder-derived model, the supersymmetry between the charge 
and the down-spins is broken, resulting in a pronounced temperature-dependence 
(Figs.~\ref{fig:slices5}a-e), in strong contrast to the persistent 
$\delta$-function nature of the up-spin response (Figs.~\ref{fig:slices5}f-j). 
We remark here that nothing close to a $\delta$-function response at $q_h = 0$ 
has been obtained in studies of cuprate $t$-$J$ chains ($t = 3J$).
\cite{kim1996,kim1997,kim2006} 

\bigskip
\noindent
{\bf{Discussion}}

\vspace{4pt}

The fractionalization of collective excitations into underlying fermionic and/or 
bosonic components can nowadays be called commonplace in models of 
correlated condensed matter. Still it is rare that such models are considered 
from the standpoint of supersymmetry, and although to date no supersymmetric 
model proposed in the condensed-matter literature has been close to an experimental 
realization, this situation is advancing in quantum optics as realized in synthetic 
quantum systems.\cite{tomka2015,tajima2021,minar2022,cai2022} Here we 
have shown how the two-leg quantum spin ladder in an 
applied magnetic field can be used not only as a ``quantum simulator'' for 
the paradigm problem of a hole propagating in a highly correlated spin 
background, but that the associated \mbox{$t$-$J$} model is supersymmetric. 
Although we do not obtain any of the spectacular consequences desired of 
supersymmetry beyond the Standard Model -- hitherto unknown superpartner 
particles at exotic energy scales reflecting broken supersymmetry -- we do 
find (i) a clear realization of an expanded supersymmetric group containing 
dual bosonic and fermionic superpartners, (ii) demonstrable consequences in 
the spectra of the zero- and one-boson sectors and 
(iii) supersymmetry-enforced kinematic constraints that become 
particularly evident in the presence of thermal fluctuations.

Away from the supersymmetric point, the doped \mbox{$t$-$J$} model still 
captures one of the most complex problems in condensed matter, specifically 
of a fractionalized ``impurity'' particle in a fractional spin environment, 
and with the quantum spin ladder this can be studied away from the 
supersymmetric wavevector (${\bf q} = {\bf 0}$). From an experimental 
standpoint, the equivalence of the ladder to a \mbox{$t$-$J$} model allows 
the energy and momentum resolution of modern INS spectrometers to be applied 
to probe fractional impurity dynamics at levels of accuracy, homogeneity, 
system size and low effective temperatures not yet accessible to other 
experimental platforms such as ultracold trapped atoms or ions. The two 
control parameters of the applied field, which modulates the spatial response 
of the environment, and the temperature, which alters the availability of 
states at different energies, allowed us to observe coherent quantum 
many-body phenomena where thermal fluctuations drive dramatic qualititative 
changes to the spectrum. 

While the ladder-to-chain mapping does not produce precisely the Heisenberg 
$t$-$J$ chain, the interaction term provides an additional ``handle'' that 
selects quasiparticle spin states preserving or breaking supersymmetry. It 
is crucial that this interaction is constrained to preserve one supersymmetry, 
and hence it cannot alter the qualitative nature of the spectra by forming 
bound states that split off from the continua. Varying the ladder parameters 
provides the insight that there exists a large family of ``ladder-derived'' 
$t$-$J$-chain models, up to and including those with strongly asymmetric 
spinon species akin to a Falicov-Kimball model, whose Heisenberg-ladder 
origin nevertheless ensures one exact pole in the spectral function and one 
supersymmetry, even if the model is no longer integrable (or has very well 
hidden integrability). 

We note again that the family of ladder-derived $t$-$J$-chain models lies 
far from the regime $t \gg J$ describing insulating spin chains with true 
electronic degrees of freedom. In the ladder-derived models, it is clear 
from our analysis that a spinon-holon description remains robust, but the 
interpretation of the spectral features is not as straightforward as in the 
large-$t$ regime, where the lower continuum edge was associated directly with 
the response of a single spinon.\cite{kim1997,kim2006} In this regime, 
the finite-temperature spectral function\cite{feiguin2010} reflects the physics 
of the spin-incoherent Luttinger liquid,\cite{cheianov2004,fiete2007} where 
the charge and spin degrees of freedom decohere at quite different temperatures 
(characterized respectively by $t$ and $J$). However, in the supersymmetric 
$t$-$J$ model, the two sectors have the same energy scale and we obtain the 
physics of the ${\bf q} = {\bf 0}$ pole that persists up to temperatures 
set by the energy scale for coherence of the ladder rung states, which is 
$J_\perp$. 

To summarize, the study of partially polarized two-leg spin-$1/2$ ladders opens
the door to deeper insight into the physics of fractional and supersymmetric 
quantum particles. The massless continuum can be used to probe spinons 
with field-controlled incommensurability and the massive continuum to probe 
supersymmetric holon-spinon fractionalization dynamics, both with interaction 
control, sector separation and, unless supersymmetry-forbidden, many-body 
thermal renormalization effects extending to energies far in excess of $T$. 
Where supersymmetry does forbid thermal broadening, it emerges as another 
candidate for the protection of quantum mechanical information from 
decoherence. Supersymmetry therefore continues to be a valuable organizing 
principle throughout physics: its application to the fermionic and bosonic 
fractions found in correlated condensed matter is in its infancy, and here 
we set in motion its experimental investigation. 

\bigskip
\noindent
{\bf\large{Methods}}

\vspace{4pt}

\noindent
{\bf{Materials.}}
Large and high-quality single crystals of both BPCB and BPCC were grown by 
a method of slow solvent evaporation.\cite{patyal1990} Further details are 
provided in Supplementary Note 1.\cite{SI} The two materials are isostructural, 
with monoclinic space group P$2_1$/c, and their low-temperature structures have 
been determined by neutron diffraction (Supplementary Note 1).\cite{ward2013,ward2017} 
The crystal symmetry mandates two types of ladder with identical interaction parameters 
but two different orientations. The ladder legs are aligned along the crystallographic $a$ 
axis and the rung alignment vectors are ${\bf r}_\pm = (0.3910, \pm 0.1625, 0.4810)$ 
for BPCB and ${\bf r}_\pm = (0.3822, \pm 0.1730, 0.4866)$ for BPCC, expressed 
in fractions of the unit cell. For both INS experiments, several crystals were 
co-aligned within 1$^{\circ}$, aided by neutron diffraction on the MORPHEUS 
instrument at the Swiss Spallation Neutron Source, SINQ, at the Paul Scherrer 
Institute (PSI), and glued onto a gold-coated aluminum 
holder. The mount for the BPCB experiment consisted of six co-aligned crystals 
with a total mass of $1.4\units{g}$. The mount of seven BPCC crystals with 
total mass $3.535\units{g}$ is shown in Supplementary Fig.~1.\cite{SI} 

\vspace{4pt}
\noindent
{\bf {Inelastic neutron scattering measurements.}}
The BPCB experiment was performed at the triple-axis spectrometer THALES 
at the Institut Laue-Langevin (ILL, Grenoble, France), which has a 
15\units{T} vertical-field magnet. The final neutron momentum was set to 
1.3\units{\AA$^{-1}$} by a PG(002) monochromator with a Be filter, which 
allowed a resolution of 0.16\units{meV} in energy and 0.14\units{r.l.u.} 
in momentum along $a^*$, and the measurement temperature was 
50\units{mK}. Further details are provided in Supplementary Note 2.\cite{SI} 

The BPCC experiment was conducted at the time-of-flight spectrometer 
LET\cite{bewley2011} at the ISIS Neutron and Muon Source at the Rutherford 
Appleton Laboratory (Chilton, United Kingdom), which has a $9\units{T}$ 
vertical magnet. The incoming neutron energy was set to $2.5\units{meV}$ 
and the sample was rotated in $1^\circ$ steps, spanning a total of 
$103^\circ$, around its $c$ axis, which was set parallel to the field. 
The energy resolution at $q_h = 0$ varied from 37\units{$\mu$eV} 
at the elastic line to 28\units{$\mu$eV} at an upper energy transfer 
of 0.9\units{meV}, and at $q_h = - 1$ from 56\units{$\mu$eV} to 
34\units{$\mu$eV}. For the purposes of analysing the $\delta$-function 
peaks in the scattered intensity, we note that the energy resolution at 
0.3\units{meV} for $q_h = - 1$ was 52\units{$\mu$eV} in both $q_\perp$ 
sectors. The momentum resolution was approximately 0.05\units{r.l.u.} 
along $a^*$. Scattering intensities were corrected using \textsc{Mantid},
\cite{arnold2014} binned into $S(\mathbf{Q},\omega)$ datasets\cite{ward2017} 
and integrated using \textsc{Horace}.\cite{ewings2016} Our raw scattering 
intensity data are shown in Supplementary Fig.~2 as part of the description 
of our data analysis and background-subtraction methodology in Supplementary 
Note 2.\cite{SI} 

\vspace{4pt}
\noindent
{\bf{Ladder and Chain MPS.}}
To calculate the spectral function of the two-leg 
ladder, and of different $t$-$J$-chain models, both at zero and at finite 
temperature, we use MPS methods in real and imaginary time.\cite{daley2004,
white2004} We compute the two-time real-space correlation function 
\begin{equation}
C(x_0,x_1;0,t) = \average{B^{\alpha}_{x_1}(t) B^{\gamma}_{x_0} (0)}_\beta
\nonumber
\end{equation}
at zero temperature, or at inverse temperature $\beta$, where $B^\alpha$ 
represents either the spin operators in the ladder (with $\alpha = \pm,z$) 
or the fermionic operator in the \mbox{$t$-$J$} model. In space, the 
correlation function is centred at rung $x_0 = L/2$ of the ladder or site 
$x_0 = L/2$ of the chain, where $L$ is the system length. In time, we 
performed real- and imaginary-time evolution using the time-evolving block 
decimation (TEBD) method in our ladder calculations and the tMPS method for 
the chain, with second- or fourth-order Trotter-Suzuki decomposition and 
purification of the density matrix at finite temperature.\cite{barthel2013} 
The codes for our ladder calculations were developed in-house and those for 
the $t$-$J$ chain were developed based on the ITensor library.\cite{itensor} 

The MPS spectral functions at different temperatures and applied magnetic 
fields were obtained from $C(x_0,x_1;0,t)$ by a discrete Fourier 
transformation to frequency-momentum space. This process conventionally 
uses a Gaussian filter function, $M(x,t) = e^{-(x / x_{\rm w})^2} 
e^{-(t / t_{\rm w})^2}$, where the filter widths in space and time, 
$x_{\rm w}$ and $t_{\rm w}$, can be used to remove finite-size effects.
\cite{kestin2019} We used our MPS spectral functions for two purposes. To 
identify the qualitative and quantitative properties of our ladder and chain 
spectra, we maximized the system length and evolution time, the latter by 
implementing linear prediction,\cite{white2008} to optimize the resolution in 
momentum and frequency. For comparison with experiment, we included the 
ladder orientations of the real material, weighted the numerical structure 
factors by the $\mathbf{Q}$-dependent scattering terms, integrated over 
${\bf Q}$-space in the same way as our treatment of the experimental data 
and selected appropriate filter parameters. Explicit technical details for 
achieving each task, meaning system sizes, time steps, filter widths and 
truncation errors, are listed in Supplementary Notes 5 and 6.\cite{SI}

\bibliography{bpcbpc}

\begin{thebibliography}{58}%
\makeatletter
\providecommand \@ifxundefined [1]{%
 \@ifx{#1\undefined}
}%
\providecommand \@ifnum [1]{%
 \ifnum #1\expandafter \@firstoftwo
 \else \expandafter \@secondoftwo
 \fi
}%
\providecommand \@ifx [1]{%
 \ifx #1\expandafter \@firstoftwo
 \else \expandafter \@secondoftwo
 \fi
}%
\providecommand \natexlab [1]{#1}%
\providecommand \enquote  [1]{``#1''}%
\providecommand \bibnamefont  [1]{#1}%
\providecommand \bibfnamefont [1]{#1}%
\providecommand \citenamefont [1]{#1}%
\providecommand \href@noop [0]{\@secondoftwo}%
\providecommand \href [0]{\begingroup \@sanitize@url \@href}%
\providecommand \@href[1]{\@@startlink{#1}\@@href}%
\providecommand \@@href[1]{\endgroup#1\@@endlink}%
\providecommand \@sanitize@url [0]{\catcode `\\12\catcode `\$12\catcode
  `\&12\catcode `\#12\catcode `\^12\catcode `\_12\catcode `\%12\relax}%
\providecommand \@@startlink[1]{}%
\providecommand \@@endlink[0]{}%
\providecommand \url  [0]{\begingroup\@sanitize@url \@url }%
\providecommand \@url [1]{\endgroup\@href {#1}{\urlprefix }}%
\providecommand \urlprefix  [0]{URL }%
\providecommand \Eprint [0]{\href }%
\providecommand \doibase [0]{https://doi.org/}%
\providecommand \selectlanguage [0]{\@gobble}%
\providecommand \bibinfo  [0]{\@secondoftwo}%
\providecommand \bibfield  [0]{\@secondoftwo}%
\providecommand \translation [1]{[#1]}%
\providecommand \BibitemOpen [0]{}%
\providecommand \bibitemStop [0]{}%
\providecommand \bibitemNoStop [0]{.\EOS\space}%
\providecommand \EOS [0]{\spacefactor3000\relax}%
\providecommand \BibitemShut  [1]{\csname bibitem#1\endcsname}%
\let\auto@bib@innerbib\@empty
\bibitem [{Ton()}]{Tong}%
  \BibitemOpen
  \bibinfo {title} {{D. Tong, Lectures on Supersymmetric Quantum Mechanics,
  https://www.damtp.cam.ac.uk/user/tong/ susyqm.html}}\BibitemShut {NoStop}%
\bibitem [{\citenamefont {Bertolini}(2024)}]{Bertolini}%
  \BibitemOpen
\bibfield  {title} {  }\bibfield  {author} {\bibinfo {author} {\bibfnamefont
  {M.}~\bibnamefont {Bertolini}},\ }\href {https://doi.org/10.1142/14026}
  {\emph {\bibinfo {title} {{Supersymmetry: From the Basics to Exact Results in
  Gauge Theories}}}}\ (\bibinfo  {publisher} {World Scientific},\ \bibinfo
  {address} {Singapore},\ \bibinfo {year} {2024})\BibitemShut {NoStop}%
\bibitem [{\citenamefont {Iachello}(1980)}]{iachello1980}%
  \BibitemOpen
  \bibfield  {author} {\bibinfo {author} {\bibfnamefont {F.}~\bibnamefont
  {Iachello}},\ }\bibfield  {title} {\bibinfo {title} {{Dynamical
  Supersymmetries in Nuclei}},\ }\href
  {https://doi.org/10.1103/PhysRevLett.44.772} {\bibfield  {journal} {\bibinfo
  {journal} {Phys. Rev. Lett.}\ }\textbf {\bibinfo {volume} {44}},\ \bibinfo
  {pages} {772} (\bibinfo {year} {1980})}\BibitemShut {NoStop}%
\bibitem [{\citenamefont {Bijker}(2010)}]{bijker2010}%
  \BibitemOpen
  \bibfield  {author} {\bibinfo {author} {\bibfnamefont {R.}~\bibnamefont
  {Bijker}},\ }\bibfield  {title} {\bibinfo {title} {{Supersymmetry in nuclear
  physics}},\ }\href {https://doi.org/10.1088/1742-6596/237/1/012005}
  {\bibfield  {journal} {\bibinfo  {journal} {J. Phys. Conf. Series}\ }\textbf
  {\bibinfo {volume} {237}},\ \bibinfo {pages} {012005} (\bibinfo {year}
  {2010})}\BibitemShut {NoStop}%
\bibitem [{\citenamefont {{Zyla, P. A. {\it et al}. (Particle Data
  Group)}}(2020)}]{mssm}%
  \BibitemOpen
  \bibfield  {author} {\bibinfo {author} {\bibnamefont {{Zyla, P. A. {\it et
  al}. (Particle Data Group)}}},\ }\bibfield  {title} {\bibinfo {title}
  {{Review of Particle Physics Ch.~88, Supersymmetry Part I (Theory)}},\ }\href
  {https://doi.org/10.1093/ptep/ptaa104} {\bibfield  {journal} {\bibinfo
  {journal} {Prog. Theor. Exp. Phys.}\ }\textbf {\bibinfo {volume} {2020}},\
  \bibinfo {pages} {083C01} (\bibinfo {year} {2020})}\BibitemShut {NoStop}%
\bibitem [{\citenamefont {Fendley}\ \emph {et~al.}(2003)\citenamefont
  {Fendley}, \citenamefont {Schoutens},\ and\ \citenamefont {{de
  Boer}}}]{fendley2003}%
  \BibitemOpen
  \bibfield  {author} {\bibinfo {author} {\bibfnamefont {P.}~\bibnamefont
  {Fendley}}, \bibinfo {author} {\bibfnamefont {K.}~\bibnamefont {Schoutens}},\
  and\ \bibinfo {author} {\bibfnamefont {J.}~\bibnamefont {{de Boer}}},\
  }\bibfield  {title} {\bibinfo {title} {{Lattice Models with ${\cal N} = 2$
  Supersymmetry}},\ }\href {https://doi.org/10.1103/PhysRevLett.90.120402}
  {\bibfield  {journal} {\bibinfo  {journal} {Phys. Rev. Lett.}\ }\textbf
  {\bibinfo {volume} {90}},\ \bibinfo {pages} {120402} (\bibinfo {year}
  {2003})}\BibitemShut {NoStop}%
\bibitem [{\citenamefont {Hagendorf}(2013)}]{hagendorf2013}%
  \BibitemOpen
  \bibfield  {author} {\bibinfo {author} {\bibfnamefont {C.}~\bibnamefont
  {Hagendorf}},\ }\bibfield  {title} {\bibinfo {title} {{Spin Chains with
  Dynamical Lattice Supersymmetry}},\ }\href
  {https://doi.org/10.1007/s10955-013-0709-9} {\bibfield  {journal} {\bibinfo
  {journal} {J. Stat. Phys.}\ }\textbf {\bibinfo {volume} {150}},\ \bibinfo
  {pages} {609} (\bibinfo {year} {2013})}\BibitemShut {NoStop}%
\bibitem [{\citenamefont {Friedan}\ \emph {et~al.}(1985)\citenamefont
  {Friedan}, \citenamefont {Qiu}, ,\ and\ \citenamefont
  {Shenker}}]{friedan1985}%
  \BibitemOpen
  \bibfield  {author} {\bibinfo {author} {\bibfnamefont {D.}~\bibnamefont
  {Friedan}}, \bibinfo {author} {\bibfnamefont {Z.-A.}\ \bibnamefont {Qiu}}, ,\
  and\ \bibinfo {author} {\bibfnamefont {S.~H.}\ \bibnamefont {Shenker}},\
  }\bibfield  {title} {\bibinfo {title} {{Superconformal invariance in two
  dimensions and the tricritical Ising model}},\ }\href
  {https://doi.org/10.1016/0370-2693(85)90819-6} {\bibfield  {journal}
  {\bibinfo  {journal} {Phys. Lett. B}\ }\textbf {\bibinfo {volume} {151}},\
  \bibinfo {pages} {37} (\bibinfo {year} {1985})}\BibitemShut {NoStop}%
\bibitem [{\citenamefont {Bauer}\ \emph {et~al.}(2013)\citenamefont {Bauer},
  \citenamefont {Huijse}, \citenamefont {Berg}, \citenamefont {Troyer},\ and\
  \citenamefont {Schoutens}}]{bauer2013}%
  \BibitemOpen
  \bibfield  {author} {\bibinfo {author} {\bibfnamefont {B.}~\bibnamefont
  {Bauer}}, \bibinfo {author} {\bibfnamefont {L.}~\bibnamefont {Huijse}},
  \bibinfo {author} {\bibfnamefont {E.}~\bibnamefont {Berg}}, \bibinfo {author}
  {\bibfnamefont {M.}~\bibnamefont {Troyer}},\ and\ \bibinfo {author}
  {\bibfnamefont {K.}~\bibnamefont {Schoutens}},\ }\bibfield  {title} {\bibinfo
  {title} {{Supersymmetric multicritical point in a model of lattice
  fermions}},\ }\href {https://doi.org/10.1103/PhysRevLett.87.165145}
  {\bibfield  {journal} {\bibinfo  {journal} {Phys. Rev. B}\ }\textbf {\bibinfo
  {volume} {87}},\ \bibinfo {pages} {165145} (\bibinfo {year}
  {2013})}\BibitemShut {NoStop}%
\bibitem [{\citenamefont {Huijse}\ \emph {et~al.}(2015)\citenamefont {Huijse},
  \citenamefont {Bauer},\ and\ \citenamefont {Berg}}]{huijse2015}%
  \BibitemOpen
  \bibfield  {author} {\bibinfo {author} {\bibfnamefont {L.}~\bibnamefont
  {Huijse}}, \bibinfo {author} {\bibfnamefont {B.}~\bibnamefont {Bauer}},\ and\
  \bibinfo {author} {\bibfnamefont {E.}~\bibnamefont {Berg}},\ }\bibfield
  {title} {\bibinfo {title} {{Emergent Supersymmetry at the
  Ising-Berezinskii-Kosterlitz-Thouless Multicritical Point}},\ }\href
  {https://doi.org/10.1103/PhysRevLett.114.090404} {\bibfield  {journal}
  {\bibinfo  {journal} {Phys. Rev. Lett.}\ }\textbf {\bibinfo {volume} {114}},\
  \bibinfo {pages} {090404} (\bibinfo {year} {2015})}\BibitemShut {NoStop}%
\bibitem [{\citenamefont {Grover}\ \emph {et~al.}(2014)\citenamefont {Grover},
  \citenamefont {Sheng},\ and\ \citenamefont {Vishwanath}}]{grover2014}%
  \BibitemOpen
  \bibfield  {author} {\bibinfo {author} {\bibfnamefont {T.}~\bibnamefont
  {Grover}}, \bibinfo {author} {\bibfnamefont {D.~N.}\ \bibnamefont {Sheng}},\
  and\ \bibinfo {author} {\bibfnamefont {A.}~\bibnamefont {Vishwanath}},\
  }\bibfield  {title} {\bibinfo {title} {{Emergent Space-Time Supersymmetry at
  the Boundary of a Topological Phase}},\ }\href
  {https://doi.org/10.1126/science.1248253} {\bibfield  {journal} {\bibinfo
  {journal} {Science}\ }\textbf {\bibinfo {volume} {344}},\ \bibinfo {pages}
  {280} (\bibinfo {year} {2014})}\BibitemShut {NoStop}%
\bibitem [{\citenamefont {Tomka}\ \emph {et~al.}(2015)\citenamefont {Tomka},
  \citenamefont {Pletyukhov},\ and\ \citenamefont {Gritsev}}]{tomka2015}%
  \BibitemOpen
  \bibfield  {author} {\bibinfo {author} {\bibfnamefont {M.}~\bibnamefont
  {Tomka}}, \bibinfo {author} {\bibfnamefont {M.}~\bibnamefont {Pletyukhov}},\
  and\ \bibinfo {author} {\bibfnamefont {V.}~\bibnamefont {Gritsev}},\
  }\bibfield  {title} {\bibinfo {title} {{Supersymmetry in quantum optics and
  in spin-orbit coupled systems}},\ }\href {https://doi.org/10.1038/srep13097}
  {\bibfield  {journal} {\bibinfo  {journal} {Sci. Rep.}\ }\textbf {\bibinfo
  {volume} {5}},\ \bibinfo {pages} {13097} (\bibinfo {year}
  {2015})}\BibitemShut {NoStop}%
\bibitem [{\citenamefont {Tajima}\ \emph {et~al.}(2021)\citenamefont {Tajima},
  \citenamefont {Hidaka},\ and\ \citenamefont {Satow}}]{tajima2021}%
  \BibitemOpen
  \bibfield  {author} {\bibinfo {author} {\bibfnamefont {H.}~\bibnamefont
  {Tajima}}, \bibinfo {author} {\bibfnamefont {Y.}~\bibnamefont {Hidaka}},\
  and\ \bibinfo {author} {\bibfnamefont {D.}~\bibnamefont {Satow}},\ }\bibfield
   {title} {\bibinfo {title} {{Goldstino spectrum in an ultracold Bose-Fermi
  mixture with explicitly broken supersymmetry}},\ }\href
  {https://doi.org/10.1103/PhysRevResearch.3.013035} {\bibfield  {journal}
  {\bibinfo  {journal} {Phys. Rev. Res.}\ }\textbf {\bibinfo {volume} {3}},\
  \bibinfo {pages} {013035} (\bibinfo {year} {2021})}\BibitemShut {NoStop}%
\bibitem [{\citenamefont {Minář}\ \emph {et~al.}(2022)\citenamefont
  {Minář}, \citenamefont {van Voorden},\ and\ \citenamefont
  {Schoutens}}]{minar2022}%
  \BibitemOpen
  \bibfield  {author} {\bibinfo {author} {\bibfnamefont {J.}~\bibnamefont
  {Minář}}, \bibinfo {author} {\bibfnamefont {B.}~\bibnamefont {van
  Voorden}},\ and\ \bibinfo {author} {\bibfnamefont {K.}~\bibnamefont
  {Schoutens}},\ }\bibfield  {title} {\bibinfo {title} {{Kink Dynamics and
  Quantum Simulation of Supersymmetric Lattice Hamiltonians}},\ }\href
  {https://doi.org/10.1103/PhysRevLett.128.050504} {\bibfield  {journal}
  {\bibinfo  {journal} {Phys. Rev. Lett.}\ }\textbf {\bibinfo {volume} {128}},\
  \bibinfo {pages} {050504} (\bibinfo {year} {2022})}\BibitemShut {NoStop}%
\bibitem [{\citenamefont {Cai}\ \emph {et~al.}(2022)\citenamefont {Cai},
  \citenamefont {Wu}, \citenamefont {Mei}, \citenamefont {Zhao}, \citenamefont
  {Jiang}, \citenamefont {Yao}, \citenamefont {He}, \citenamefont {Zhou},\ and\
  \citenamefont {Duan}}]{cai2022}%
  \BibitemOpen
  \bibfield  {author} {\bibinfo {author} {\bibfnamefont {M.-L.}\ \bibnamefont
  {Cai}}, \bibinfo {author} {\bibfnamefont {Y.-K.}\ \bibnamefont {Wu}},
  \bibinfo {author} {\bibfnamefont {Q.-X.}\ \bibnamefont {Mei}}, \bibinfo
  {author} {\bibfnamefont {W.-D.}\ \bibnamefont {Zhao}}, \bibinfo {author}
  {\bibfnamefont {Y.}~\bibnamefont {Jiang}}, \bibinfo {author} {\bibfnamefont
  {L.}~\bibnamefont {Yao}}, \bibinfo {author} {\bibfnamefont {L.}~\bibnamefont
  {He}}, \bibinfo {author} {\bibfnamefont {Z.-C.}\ \bibnamefont {Zhou}},\ and\
  \bibinfo {author} {\bibfnamefont {L.-M.}\ \bibnamefont {Duan}},\ }\bibfield
  {title} {\bibinfo {title} {Observation of supersymmetry and its spontaneous
  breaking in a trapped ion quantum simulator},\ }\href
  {https://doi.org/10.1038/s41467-022-31058-0} {\bibfield  {journal} {\bibinfo
  {journal} {Nature Commun.}\ }\textbf {\bibinfo {volume} {13}},\ \bibinfo
  {pages} {3412} (\bibinfo {year} {2022})}\BibitemShut {NoStop}%
\bibitem [{\citenamefont {Imada}\ \emph {et~al.}(1998)\citenamefont {Imada},
  \citenamefont {Fujimori},\ and\ \citenamefont {Tokura}}]{imada1998}%
  \BibitemOpen
  \bibfield  {author} {\bibinfo {author} {\bibfnamefont {M.}~\bibnamefont
  {Imada}}, \bibinfo {author} {\bibfnamefont {A.}~\bibnamefont {Fujimori}},\
  and\ \bibinfo {author} {\bibfnamefont {Y.}~\bibnamefont {Tokura}},\
  }\bibfield  {title} {\bibinfo {title} {{Metal-insulator transitions}},\
  }\href {https://doi.org/10.1103/RevModPhys.70.1039} {\bibfield  {journal}
  {\bibinfo  {journal} {Rev. Mod. Phys.}\ }\textbf {\bibinfo {volume} {70}},\
  \bibinfo {pages} {1039} (\bibinfo {year} {1998})}\BibitemShut {NoStop}%
\bibitem [{\citenamefont {Lee}\ \emph {et~al.}(2006)\citenamefont {Lee},
  \citenamefont {Nagaosa},\ and\ \citenamefont {Wen}}]{lee2006}%
  \BibitemOpen
  \bibfield  {author} {\bibinfo {author} {\bibfnamefont {P.~A.}\ \bibnamefont
  {Lee}}, \bibinfo {author} {\bibfnamefont {N.}~\bibnamefont {Nagaosa}},\ and\
  \bibinfo {author} {\bibfnamefont {X.-G.}\ \bibnamefont {Wen}},\ }\bibfield
  {title} {\bibinfo {title} {{Doping a Mott insulator: Physics of
  high-temperature superconductivity}},\ }\href
  {https://doi.org/10.1103/RevModPhys.78.17} {\bibfield  {journal} {\bibinfo
  {journal} {Rev. Mod. Phys.}\ }\textbf {\bibinfo {volume} {78}},\ \bibinfo
  {pages} {17} (\bibinfo {year} {2006})}\BibitemShut {NoStop}%
\bibitem [{\citenamefont {Lieb}\ and\ \citenamefont {Wu}(1968)}]{lieb1968}%
  \BibitemOpen
  \bibfield  {author} {\bibinfo {author} {\bibfnamefont {E.~H.}\ \bibnamefont
  {Lieb}}\ and\ \bibinfo {author} {\bibfnamefont {F.~Y.}\ \bibnamefont {Wu}},\
  }\bibfield  {title} {\bibinfo {title} {{Absence of Mott Transition in an
  Exact Solution of the Short-Range, One-Band Model in One Dimension}},\ }\href
  {https://doi.org/10.1103/PhysRevLett.20.1445} {\bibfield  {journal} {\bibinfo
   {journal} {Phys. Rev. Lett.}\ }\textbf {\bibinfo {volume} {20}},\ \bibinfo
  {pages} {1445} (\bibinfo {year} {1968})}\BibitemShut {NoStop}%
\bibitem [{\citenamefont {Bares}\ and\ \citenamefont
  {Blatter}(1990)}]{bares1990}%
  \BibitemOpen
  \bibfield  {author} {\bibinfo {author} {\bibfnamefont {P.~A.}\ \bibnamefont
  {Bares}}\ and\ \bibinfo {author} {\bibfnamefont {G.}~\bibnamefont
  {Blatter}},\ }\bibfield  {title} {\bibinfo {title} {{Supersymmetric $t$-$J$
  model in one dimension: Separation of spin and charge}},\ }\href
  {https://doi.org/10.1103/PhysRevLett.64.2567} {\bibfield  {journal} {\bibinfo
   {journal} {Phys. Rev. Lett.}\ }\textbf {\bibinfo {volume} {64}},\ \bibinfo
  {pages} {2567} (\bibinfo {year} {1990})}\BibitemShut {NoStop}%
\bibitem [{\citenamefont {Bares}\ \emph {et~al.}(1991)\citenamefont {Bares},
  \citenamefont {Blatter},\ and\ \citenamefont {Ogata}}]{bares1991}%
  \BibitemOpen
  \bibfield  {author} {\bibinfo {author} {\bibfnamefont {P.-A.}\ \bibnamefont
  {Bares}}, \bibinfo {author} {\bibfnamefont {G.}~\bibnamefont {Blatter}},\
  and\ \bibinfo {author} {\bibfnamefont {M.}~\bibnamefont {Ogata}},\ }\bibfield
   {title} {\bibinfo {title} {{Exact solution of the $t$-$J$ model in one
  dimension at $2t = \pm J$: Ground state and excitation spectrum}},\ }\href
  {https://doi.org/10.1103/PhysRevB.44.130} {\bibfield  {journal} {\bibinfo
  {journal} {Phys. Rev. B}\ }\textbf {\bibinfo {volume} {44}},\ \bibinfo
  {pages} {130} (\bibinfo {year} {1991})}\BibitemShut {NoStop}%
\bibitem [{\citenamefont {Essler}\ and\ \citenamefont
  {Korepin}(1992)}]{essler1992}%
  \BibitemOpen
  \bibfield  {author} {\bibinfo {author} {\bibfnamefont {F.~H.~L.}\
  \bibnamefont {Essler}}\ and\ \bibinfo {author} {\bibfnamefont {V.~E.}\
  \bibnamefont {Korepin}},\ }\bibfield  {title} {\bibinfo {title} {{Higher
  conservation laws and algebraic Bethe Ans\"atze for the supersymmetric
  $t$-$J$ model}},\ }\href {https://doi.org/10.1103/PhysRevB.46.9147}
  {\bibfield  {journal} {\bibinfo  {journal} {Phys. Rev. B}\ }\textbf {\bibinfo
  {volume} {46}},\ \bibinfo {pages} {9147} (\bibinfo {year}
  {1992})}\BibitemShut {NoStop}%
\bibitem [{\citenamefont {Segovia}\ \emph {et~al.}(1999)\citenamefont
  {Segovia}, \citenamefont {Purdie}, \citenamefont {Hengsberger},\ and\
  \citenamefont {Baer}}]{segovia1999}%
  \BibitemOpen
  \bibfield  {author} {\bibinfo {author} {\bibfnamefont {P.}~\bibnamefont
  {Segovia}}, \bibinfo {author} {\bibfnamefont {D.}~\bibnamefont {Purdie}},
  \bibinfo {author} {\bibfnamefont {M.}~\bibnamefont {Hengsberger}},\ and\
  \bibinfo {author} {\bibfnamefont {Y.}~\bibnamefont {Baer}},\ }\bibfield
  {title} {\bibinfo {title} {Observation of spin and charge collective modes in
  one-dimensional metallic chains},\ }\href {https://doi.org/10.1038/990052}
  {\bibfield  {journal} {\bibinfo  {journal} {Nature}\ }\textbf {\bibinfo
  {volume} {402}},\ \bibinfo {pages} {504} (\bibinfo {year}
  {1999})}\BibitemShut {NoStop}%
\bibitem [{\citenamefont {Claessen}\ \emph {et~al.}(2002)\citenamefont
  {Claessen}, \citenamefont {Sing}, \citenamefont {Schwingenschl\"ogl},
  \citenamefont {Dressel},\ and\ \citenamefont {Jacobsen}}]{claessen2002}%
  \BibitemOpen
  \bibfield  {author} {\bibinfo {author} {\bibfnamefont {R.}~\bibnamefont
  {Claessen}}, \bibinfo {author} {\bibfnamefont {M.}~\bibnamefont {Sing}},
  \bibinfo {author} {\bibfnamefont {U.}~\bibnamefont {Schwingenschl\"ogl}},
  \bibinfo {author} {\bibfnamefont {M.}~\bibnamefont {Dressel}},\ and\ \bibinfo
  {author} {\bibfnamefont {C.~S.}\ \bibnamefont {Jacobsen}},\ }\bibfield
  {title} {\bibinfo {title} {{Spectroscopic Signatures of Spin-Charge
  Separation in the Quasi-One-Dimensional Organic Conductor TTF-TCNQ}},\ }\href
  {https://doi.org/10.1103/PhysRevLett.88.096402} {\bibfield  {journal}
  {\bibinfo  {journal} {Phys. Rev. Lett.}\ }\textbf {\bibinfo {volume} {88}},\
  \bibinfo {pages} {096402} (\bibinfo {year} {2002})}\BibitemShut {NoStop}%
\bibitem [{\citenamefont {Kim}\ \emph {et~al.}(1996)\citenamefont {Kim},
  \citenamefont {Matsuura}, \citenamefont {Shen}, \citenamefont {Motoyama},
  \citenamefont {Eisaki}, \citenamefont {Uchida}, \citenamefont {Tohyama},\
  and\ \citenamefont {Maekawa}}]{kim1996}%
  \BibitemOpen
  \bibfield  {author} {\bibinfo {author} {\bibfnamefont {C.}~\bibnamefont
  {Kim}}, \bibinfo {author} {\bibfnamefont {A.~Y.}\ \bibnamefont {Matsuura}},
  \bibinfo {author} {\bibfnamefont {Z.-X.}\ \bibnamefont {Shen}}, \bibinfo
  {author} {\bibfnamefont {N.}~\bibnamefont {Motoyama}}, \bibinfo {author}
  {\bibfnamefont {H.}~\bibnamefont {Eisaki}}, \bibinfo {author} {\bibfnamefont
  {S.}~\bibnamefont {Uchida}}, \bibinfo {author} {\bibfnamefont
  {T.}~\bibnamefont {Tohyama}},\ and\ \bibinfo {author} {\bibfnamefont
  {S.}~\bibnamefont {Maekawa}},\ }\bibfield  {title} {\bibinfo {title}
  {{Observation of Spin-Charge Separation in One-Dimensional SrCuO$_2$}},\
  }\href {https://doi.org/10.1103/PhysRevLett.77.4054} {\bibfield  {journal}
  {\bibinfo  {journal} {Phys. Rev. Lett.}\ }\textbf {\bibinfo {volume} {77}},\
  \bibinfo {pages} {4054} (\bibinfo {year} {1996})}\BibitemShut {NoStop}%
\bibitem [{\citenamefont {Kim}\ \emph {et~al.}(1997)\citenamefont {Kim},
  \citenamefont {Shen}, \citenamefont {Motoyama}, \citenamefont {Eisaki},
  \citenamefont {Uchida}, \citenamefont {Tohyama},\ and\ \citenamefont
  {Maekawa}}]{kim1997}%
  \BibitemOpen
  \bibfield  {author} {\bibinfo {author} {\bibfnamefont {C.}~\bibnamefont
  {Kim}}, \bibinfo {author} {\bibfnamefont {Z.-X.}\ \bibnamefont {Shen}},
  \bibinfo {author} {\bibfnamefont {N.}~\bibnamefont {Motoyama}}, \bibinfo
  {author} {\bibfnamefont {H.}~\bibnamefont {Eisaki}}, \bibinfo {author}
  {\bibfnamefont {S.}~\bibnamefont {Uchida}}, \bibinfo {author} {\bibfnamefont
  {T.}~\bibnamefont {Tohyama}},\ and\ \bibinfo {author} {\bibfnamefont
  {S.}~\bibnamefont {Maekawa}},\ }\bibfield  {title} {\bibinfo {title}
  {Separation of spin and charge excitations in one-dimensional {SrCuO$_2$}},\
  }\href {https://doi.org/10.1103/PhysRevB.56.15589} {\bibfield  {journal}
  {\bibinfo  {journal} {Phys. Rev. B}\ }\textbf {\bibinfo {volume} {56}},\
  \bibinfo {pages} {15589} (\bibinfo {year} {1997})}\BibitemShut {NoStop}%
\bibitem [{\citenamefont {Kim}\ \emph {et~al.}(2006)\citenamefont {Kim},
  \citenamefont {Koh}, \citenamefont {Rotenberg}, \citenamefont {Oh},
  \citenamefont {Eisaki}, \citenamefont {Motoyama}, \citenamefont {Uchida},
  \citenamefont {Tohyama}, \citenamefont {Maekawa}, \citenamefont {Shen},\ and\
  \citenamefont {Kim}}]{kim2006}%
  \BibitemOpen
  \bibfield  {author} {\bibinfo {author} {\bibfnamefont {B.~J.}\ \bibnamefont
  {Kim}}, \bibinfo {author} {\bibfnamefont {H.}~\bibnamefont {Koh}}, \bibinfo
  {author} {\bibfnamefont {E.}~\bibnamefont {Rotenberg}}, \bibinfo {author}
  {\bibfnamefont {S.-J.}\ \bibnamefont {Oh}}, \bibinfo {author} {\bibfnamefont
  {H.}~\bibnamefont {Eisaki}}, \bibinfo {author} {\bibfnamefont
  {N.}~\bibnamefont {Motoyama}}, \bibinfo {author} {\bibfnamefont
  {S.}~\bibnamefont {Uchida}}, \bibinfo {author} {\bibfnamefont
  {T.}~\bibnamefont {Tohyama}}, \bibinfo {author} {\bibfnamefont
  {S.}~\bibnamefont {Maekawa}}, \bibinfo {author} {\bibfnamefont {Z.-X.}\
  \bibnamefont {Shen}},\ and\ \bibinfo {author} {\bibfnamefont
  {C.}~\bibnamefont {Kim}},\ }\bibfield  {title} {\bibinfo {title} {Distinct
  spinon and holon dispersions in photoemission spectral functions from
  one-dimensional {SrCuO$_2$}},\ }\href {https://doi.org/10.1038/nphys316}
  {\bibfield  {journal} {\bibinfo  {journal} {Nature Phys.}\ }\textbf {\bibinfo
  {volume} {2}},\ \bibinfo {pages} {397} (\bibinfo {year} {2006})}\BibitemShut
  {NoStop}%
\bibitem [{\citenamefont {Vijayan}\ \emph {et~al.}(2020)\citenamefont
  {Vijayan}, \citenamefont {Sompet}, \citenamefont {Salomon}, \citenamefont
  {Koepsell}, \citenamefont {Hirthe}, \citenamefont {Bohrdt}, \citenamefont
  {Grusdt}, \citenamefont {Bloch},\ and\ \citenamefont {Gross}}]{vijayan2020}%
  \BibitemOpen
  \bibfield  {author} {\bibinfo {author} {\bibfnamefont {J.}~\bibnamefont
  {Vijayan}}, \bibinfo {author} {\bibfnamefont {P.}~\bibnamefont {Sompet}},
  \bibinfo {author} {\bibfnamefont {G.}~\bibnamefont {Salomon}}, \bibinfo
  {author} {\bibfnamefont {J.}~\bibnamefont {Koepsell}}, \bibinfo {author}
  {\bibfnamefont {S.}~\bibnamefont {Hirthe}}, \bibinfo {author} {\bibfnamefont
  {A.}~\bibnamefont {Bohrdt}}, \bibinfo {author} {\bibfnamefont
  {F.}~\bibnamefont {Grusdt}}, \bibinfo {author} {\bibfnamefont
  {I.}~\bibnamefont {Bloch}},\ and\ \bibinfo {author} {\bibfnamefont
  {C.}~\bibnamefont {Gross}},\ }\bibfield  {title} {\bibinfo {title}
  {Time-resolved observation of spin-charge deconfinement in fermionic
  {Hubbard} chains},\ }\href {https://doi.org/10.1126/science.aay2354}
  {\bibfield  {journal} {\bibinfo  {journal} {Science}\ }\textbf {\bibinfo
  {volume} {367}},\ \bibinfo {pages} {186} (\bibinfo {year}
  {2020})}\BibitemShut {NoStop}%
\bibitem [{\citenamefont {Schlappa}\ \emph {et~al.}(2012)\citenamefont
  {Schlappa}, \citenamefont {Wohlfeld}, \citenamefont {Zhou}, \citenamefont
  {Mourigal}, \citenamefont {Haverkort}, \citenamefont {Strocov}, \citenamefont
  {Hozoi}, \citenamefont {Monney}, \citenamefont {Nishimoto}, \citenamefont
  {Singh}, \citenamefont {Revcolevschi}, \citenamefont {Caux}, \citenamefont
  {Patthey}, \citenamefont {R\o{}nnow}, \citenamefont {van~den Brink},\ and\
  \citenamefont {Schmitt}}]{schlappa2012}%
  \BibitemOpen
  \bibfield  {author} {\bibinfo {author} {\bibfnamefont {J.}~\bibnamefont
  {Schlappa}}, \bibinfo {author} {\bibfnamefont {K.}~\bibnamefont {Wohlfeld}},
  \bibinfo {author} {\bibfnamefont {K.~J.}\ \bibnamefont {Zhou}}, \bibinfo
  {author} {\bibfnamefont {M.}~\bibnamefont {Mourigal}}, \bibinfo {author}
  {\bibfnamefont {M.~W.}\ \bibnamefont {Haverkort}}, \bibinfo {author}
  {\bibfnamefont {V.~N.}\ \bibnamefont {Strocov}}, \bibinfo {author}
  {\bibfnamefont {L.}~\bibnamefont {Hozoi}}, \bibinfo {author} {\bibfnamefont
  {C.}~\bibnamefont {Monney}}, \bibinfo {author} {\bibfnamefont
  {S.}~\bibnamefont {Nishimoto}}, \bibinfo {author} {\bibfnamefont
  {S.}~\bibnamefont {Singh}}, \bibinfo {author} {\bibfnamefont
  {A.}~\bibnamefont {Revcolevschi}}, \bibinfo {author} {\bibfnamefont {J.-S.}\
  \bibnamefont {Caux}}, \bibinfo {author} {\bibfnamefont {L.}~\bibnamefont
  {Patthey}}, \bibinfo {author} {\bibfnamefont {H.~M.}\ \bibnamefont
  {R\o{}nnow}}, \bibinfo {author} {\bibfnamefont {J.}~\bibnamefont {van~den
  Brink}},\ and\ \bibinfo {author} {\bibfnamefont {T.}~\bibnamefont
  {Schmitt}},\ }\bibfield  {title} {\bibinfo {title} {{Spin-orbital separation
  in the quasi-one-dimensional Mott insulator Sr$_2$CuO$_3$}},\ }\href
  {https://doi.org/10.1038/nature10974} {\bibfield  {journal} {\bibinfo
  {journal} {Nature}\ }\textbf {\bibinfo {volume} {485}},\ \bibinfo {pages}
  {82} (\bibinfo {year} {2012})}\BibitemShut {NoStop}%
\bibitem [{\citenamefont {Klanj\v{s}ek}\ \emph {et~al.}(2008)\citenamefont
  {Klanj\v{s}ek}, \citenamefont {Mayaffre}, \citenamefont {Berthier},
  \citenamefont {Horvati\'c}, \citenamefont {Chiari}, \citenamefont
  {Piovesana}, \citenamefont {Bouillot}, \citenamefont {Kollath}, \citenamefont
  {Orignac}, \citenamefont {Citro},\ and\ \citenamefont
  {Giamarchi}}]{klanjsek2008}%
  \BibitemOpen
  \bibfield  {author} {\bibinfo {author} {\bibfnamefont {M.}~\bibnamefont
  {Klanj\v{s}ek}}, \bibinfo {author} {\bibfnamefont {H.}~\bibnamefont
  {Mayaffre}}, \bibinfo {author} {\bibfnamefont {C.}~\bibnamefont {Berthier}},
  \bibinfo {author} {\bibfnamefont {M.}~\bibnamefont {Horvati\'c}}, \bibinfo
  {author} {\bibfnamefont {B.}~\bibnamefont {Chiari}}, \bibinfo {author}
  {\bibfnamefont {O.}~\bibnamefont {Piovesana}}, \bibinfo {author}
  {\bibfnamefont {P.}~\bibnamefont {Bouillot}}, \bibinfo {author}
  {\bibfnamefont {C.}~\bibnamefont {Kollath}}, \bibinfo {author} {\bibfnamefont
  {E.}~\bibnamefont {Orignac}}, \bibinfo {author} {\bibfnamefont
  {R.}~\bibnamefont {Citro}},\ and\ \bibinfo {author} {\bibfnamefont
  {T.}~\bibnamefont {Giamarchi}},\ }\bibfield  {title} {\bibinfo {title}
  {Controlling {Luttinger} {Liquid} {Physics} in {Spin} {Ladders} under a
  {Magnetic} {Field}},\ }\href {https://doi.org/10.1103/PhysRevLett.101.137207}
  {\bibfield  {journal} {\bibinfo  {journal} {Phys. Rev. Lett.}\ }\textbf
  {\bibinfo {volume} {101}},\ \bibinfo {pages} {137207} (\bibinfo {year}
  {2008})}\BibitemShut {NoStop}%
\bibitem [{\citenamefont {Thielemann}\ \emph
  {et~al.}(2009{\natexlab{a}})\citenamefont {Thielemann}, \citenamefont
  {R\"uegg}, \citenamefont {R\o{}nnow}, \citenamefont {L\"auchli},
  \citenamefont {Caux}, \citenamefont {Normand}, \citenamefont {Biner},
  \citenamefont {Kr\"amer}, \citenamefont {G\"udel}, \citenamefont {Stahn},
  \citenamefont {Habicht}, \citenamefont {Kiefer}, \citenamefont {Boehm},
  \citenamefont {McMorrow},\ and\ \citenamefont {Mesot}}]{thielemann2009a}%
  \BibitemOpen
  \bibfield  {author} {\bibinfo {author} {\bibfnamefont {B.}~\bibnamefont
  {Thielemann}}, \bibinfo {author} {\bibfnamefont {C.}~\bibnamefont {R\"uegg}},
  \bibinfo {author} {\bibfnamefont {H.~M.}\ \bibnamefont {R\o{}nnow}}, \bibinfo
  {author} {\bibfnamefont {A.~M.}\ \bibnamefont {L\"auchli}}, \bibinfo {author}
  {\bibfnamefont {J.-S.}\ \bibnamefont {Caux}}, \bibinfo {author}
  {\bibfnamefont {B.}~\bibnamefont {Normand}}, \bibinfo {author} {\bibfnamefont
  {D.}~\bibnamefont {Biner}}, \bibinfo {author} {\bibfnamefont {K.~W.}\
  \bibnamefont {Kr\"amer}}, \bibinfo {author} {\bibfnamefont {H.-U.}\
  \bibnamefont {G\"udel}}, \bibinfo {author} {\bibfnamefont {J.}~\bibnamefont
  {Stahn}}, \bibinfo {author} {\bibfnamefont {K.}~\bibnamefont {Habicht}},
  \bibinfo {author} {\bibfnamefont {K.}~\bibnamefont {Kiefer}}, \bibinfo
  {author} {\bibfnamefont {M.}~\bibnamefont {Boehm}}, \bibinfo {author}
  {\bibfnamefont {D.~F.}\ \bibnamefont {McMorrow}},\ and\ \bibinfo {author}
  {\bibfnamefont {J.}~\bibnamefont {Mesot}},\ }\bibfield  {title} {\bibinfo
  {title} {Direct {Observation} of {Magnon} {Fractionalization} in the
  {Quantum} {Spin} {Ladder}},\ }\href
  {https://doi.org/10.1103/PhysRevLett.102.107204} {\bibfield  {journal}
  {\bibinfo  {journal} {Phys. Rev. Lett.}\ }\textbf {\bibinfo {volume} {102}},\
  \bibinfo {pages} {107204} (\bibinfo {year} {2009}{\natexlab{a}})}\BibitemShut
  {NoStop}%
\bibitem [{\citenamefont {Schmidiger}\ \emph {et~al.}(2013)\citenamefont
  {Schmidiger}, \citenamefont {Bouillot}, \citenamefont {Guidi}, \citenamefont
  {Bewley}, \citenamefont {Kollath}, \citenamefont {Giamarchi},\ and\
  \citenamefont {Zheludev}}]{schmidiger2013}%
  \BibitemOpen
  \bibfield  {author} {\bibinfo {author} {\bibfnamefont {D.}~\bibnamefont
  {Schmidiger}}, \bibinfo {author} {\bibfnamefont {P.}~\bibnamefont
  {Bouillot}}, \bibinfo {author} {\bibfnamefont {T.}~\bibnamefont {Guidi}},
  \bibinfo {author} {\bibfnamefont {R.}~\bibnamefont {Bewley}}, \bibinfo
  {author} {\bibfnamefont {C.}~\bibnamefont {Kollath}}, \bibinfo {author}
  {\bibfnamefont {T.}~\bibnamefont {Giamarchi}},\ and\ \bibinfo {author}
  {\bibfnamefont {A.}~\bibnamefont {Zheludev}},\ }\bibfield  {title} {\bibinfo
  {title} {Spectrum of a {Magnetized} {Strong}-{Leg} {Quantum} {Spin}
  {Ladder}},\ }\href {https://doi.org/10.1103/PhysRevLett.111.107202}
  {\bibfield  {journal} {\bibinfo  {journal} {Phys. Rev. Lett.}\ }\textbf
  {\bibinfo {volume} {111}},\ \bibinfo {pages} {107202} (\bibinfo {year}
  {2013})}\BibitemShut {NoStop}%
\bibitem [{\citenamefont {Ward}\ \emph {et~al.}(2017)\citenamefont {Ward},
  \citenamefont {Mena}, \citenamefont {Bouillot}, \citenamefont {Kollath},
  \citenamefont {Giamarchi}, \citenamefont {Schmidt}, \citenamefont {Normand},
  \citenamefont {Kr\"amer}, \citenamefont {Biner}, \citenamefont {Bewley},
  \citenamefont {Guidi}, \citenamefont {Boehm}, \citenamefont {McMorrow},\ and\
  \citenamefont {R\"uegg}}]{ward2017}%
  \BibitemOpen
  \bibfield  {author} {\bibinfo {author} {\bibfnamefont {S.}~\bibnamefont
  {Ward}}, \bibinfo {author} {\bibfnamefont {M.}~\bibnamefont {Mena}}, \bibinfo
  {author} {\bibfnamefont {P.}~\bibnamefont {Bouillot}}, \bibinfo {author}
  {\bibfnamefont {C.}~\bibnamefont {Kollath}}, \bibinfo {author} {\bibfnamefont
  {T.}~\bibnamefont {Giamarchi}}, \bibinfo {author} {\bibfnamefont {K.~P.}\
  \bibnamefont {Schmidt}}, \bibinfo {author} {\bibfnamefont {B.}~\bibnamefont
  {Normand}}, \bibinfo {author} {\bibfnamefont {K.~W.}\ \bibnamefont
  {Kr\"amer}}, \bibinfo {author} {\bibfnamefont {D.}~\bibnamefont {Biner}},
  \bibinfo {author} {\bibfnamefont {R.}~\bibnamefont {Bewley}}, \bibinfo
  {author} {\bibfnamefont {T.}~\bibnamefont {Guidi}}, \bibinfo {author}
  {\bibfnamefont {M.}~\bibnamefont {Boehm}}, \bibinfo {author} {\bibfnamefont
  {D.~F.}\ \bibnamefont {McMorrow}},\ and\ \bibinfo {author} {\bibfnamefont
  {C.}~\bibnamefont {R\"uegg}},\ }\bibfield  {title} {\bibinfo {title} {Bound
  {States} and {Field}-{Polarized} {Haldane} {Modes} in a {Quantum} {Spin}
  {Ladder}},\ }\href {https://doi.org/10.1103/PhysRevLett.118.177202}
  {\bibfield  {journal} {\bibinfo  {journal} {Phys. Rev. Lett.}\ }\textbf
  {\bibinfo {volume} {118}},\ \bibinfo {pages} {177202} (\bibinfo {year}
  {2017})}\BibitemShut {NoStop}%
\bibitem [{\citenamefont {Bouillot}\ \emph {et~al.}(2011)\citenamefont
  {Bouillot}, \citenamefont {Kollath}, \citenamefont {L\"auchli}, \citenamefont
  {Zvonarev}, \citenamefont {Thielemann}, \citenamefont {R\"uegg},
  \citenamefont {Orignac}, \citenamefont {Citro}, \citenamefont
  {Klanj\ifmmode~\check{s}\else \v{s}\fi{}ek}, \citenamefont {Berthier},
  \citenamefont {Horvati\ifmmode~\acute{c}\else \'{c}\fi{}},\ and\
  \citenamefont {Giamarchi}}]{bouillot2011}%
  \BibitemOpen
  \bibfield  {author} {\bibinfo {author} {\bibfnamefont {P.}~\bibnamefont
  {Bouillot}}, \bibinfo {author} {\bibfnamefont {C.}~\bibnamefont {Kollath}},
  \bibinfo {author} {\bibfnamefont {A.~M.}\ \bibnamefont {L\"auchli}}, \bibinfo
  {author} {\bibfnamefont {M.}~\bibnamefont {Zvonarev}}, \bibinfo {author}
  {\bibfnamefont {B.}~\bibnamefont {Thielemann}}, \bibinfo {author}
  {\bibfnamefont {C.}~\bibnamefont {R\"uegg}}, \bibinfo {author} {\bibfnamefont
  {E.}~\bibnamefont {Orignac}}, \bibinfo {author} {\bibfnamefont
  {R.}~\bibnamefont {Citro}}, \bibinfo {author} {\bibfnamefont
  {M.}~\bibnamefont {Klanj\ifmmode~\check{s}\else \v{s}\fi{}ek}}, \bibinfo
  {author} {\bibfnamefont {C.}~\bibnamefont {Berthier}}, \bibinfo {author}
  {\bibfnamefont {M.}~\bibnamefont {Horvati\ifmmode~\acute{c}\else
  \'{c}\fi{}}},\ and\ \bibinfo {author} {\bibfnamefont {T.}~\bibnamefont
  {Giamarchi}},\ }\bibfield  {title} {\bibinfo {title} {Statics and dynamics of
  weakly coupled antiferromagnetic spin-$\frac{1}{2}$ ladders in a magnetic
  field},\ }\href {https://doi.org/10.1103/PhysRevB.83.054407} {\bibfield
  {journal} {\bibinfo  {journal} {Phys. Rev. B}\ }\textbf {\bibinfo {volume}
  {83}},\ \bibinfo {pages} {054407} (\bibinfo {year} {2011})}\BibitemShut
  {NoStop}%
\bibitem [{\citenamefont {Ward}\ \emph {et~al.}(2013)\citenamefont {Ward},
  \citenamefont {Bouillot}, \citenamefont {Ryll}, \citenamefont {Kiefer},
  \citenamefont {Kr\"amer}, \citenamefont {R\"uegg}, \citenamefont {Kollath},\
  and\ \citenamefont {Giamarchi}}]{ward2013}%
  \BibitemOpen
  \bibfield  {author} {\bibinfo {author} {\bibfnamefont {S.}~\bibnamefont
  {Ward}}, \bibinfo {author} {\bibfnamefont {P.}~\bibnamefont {Bouillot}},
  \bibinfo {author} {\bibfnamefont {H.}~\bibnamefont {Ryll}}, \bibinfo {author}
  {\bibfnamefont {K.}~\bibnamefont {Kiefer}}, \bibinfo {author} {\bibfnamefont
  {K.~W.}\ \bibnamefont {Kr\"amer}}, \bibinfo {author} {\bibfnamefont
  {C.}~\bibnamefont {R\"uegg}}, \bibinfo {author} {\bibfnamefont
  {C.}~\bibnamefont {Kollath}},\ and\ \bibinfo {author} {\bibfnamefont
  {T.}~\bibnamefont {Giamarchi}},\ }\bibfield  {title} {\bibinfo {title} {Spin
  ladders and quantum simulators for {Tomonaga}-{Luttinger} liquids},\ }\href
  {https://doi.org/10.1088/0953-8984/25/1/014004} {\bibfield  {journal}
  {\bibinfo  {journal} {J. Phys.: Condens. Matter}\ }\textbf {\bibinfo {volume}
  {25}},\ \bibinfo {pages} {014004} (\bibinfo {year} {2013})}\BibitemShut
  {NoStop}%
\bibitem [{SI()}]{SI}%
  \BibitemOpen
  \href@noop {} {}\bibinfo {howpublished} {The Supplementary Information
  contains full details of the ladder materials BPCB and BPCC in an applied
  magnetic field, our two INS experiments, data analysis, relevant
  supersymmetry considerations and our MPS calculations for both the ladder and
  the $t$-$J$-chain models at both zero and finite temperature.}\BibitemShut
  {Stop}%
\bibitem [{\citenamefont {Thielemann}\ \emph
  {et~al.}(2009{\natexlab{b}})\citenamefont {Thielemann}, \citenamefont
  {R\"uegg}, \citenamefont {Kiefer}, \citenamefont {R\o{}nnow}, \citenamefont
  {Normand}, \citenamefont {Bouillot}, \citenamefont {Kollath}, \citenamefont
  {Orignac}, \citenamefont {Citro}, \citenamefont {Giamarchi}, \citenamefont
  {L\"auchli}, \citenamefont {Biner}, \citenamefont {Kr\"amer}, \citenamefont
  {Wolff-Fabris}, \citenamefont {Zapf}, \citenamefont {Jaime}, \citenamefont
  {Stahn}, \citenamefont {Christensen}, \citenamefont {Grenier}, \citenamefont
  {McMorrow},\ and\ \citenamefont {Mesot}}]{thielemann2009}%
  \BibitemOpen
  \bibfield  {author} {\bibinfo {author} {\bibfnamefont {B.}~\bibnamefont
  {Thielemann}}, \bibinfo {author} {\bibfnamefont {C.}~\bibnamefont {R\"uegg}},
  \bibinfo {author} {\bibfnamefont {K.}~\bibnamefont {Kiefer}}, \bibinfo
  {author} {\bibfnamefont {H.~M.}\ \bibnamefont {R\o{}nnow}}, \bibinfo {author}
  {\bibfnamefont {B.}~\bibnamefont {Normand}}, \bibinfo {author} {\bibfnamefont
  {P.}~\bibnamefont {Bouillot}}, \bibinfo {author} {\bibfnamefont
  {C.}~\bibnamefont {Kollath}}, \bibinfo {author} {\bibfnamefont
  {E.}~\bibnamefont {Orignac}}, \bibinfo {author} {\bibfnamefont
  {R.}~\bibnamefont {Citro}}, \bibinfo {author} {\bibfnamefont
  {T.}~\bibnamefont {Giamarchi}}, \bibinfo {author} {\bibfnamefont {A.~M.}\
  \bibnamefont {L\"auchli}}, \bibinfo {author} {\bibfnamefont {D.}~\bibnamefont
  {Biner}}, \bibinfo {author} {\bibfnamefont {K.~W.}\ \bibnamefont {Kr\"amer}},
  \bibinfo {author} {\bibfnamefont {F.}~\bibnamefont {Wolff-Fabris}}, \bibinfo
  {author} {\bibfnamefont {V.~S.}\ \bibnamefont {Zapf}}, \bibinfo {author}
  {\bibfnamefont {M.}~\bibnamefont {Jaime}}, \bibinfo {author} {\bibfnamefont
  {J.}~\bibnamefont {Stahn}}, \bibinfo {author} {\bibfnamefont {N.~B.}\
  \bibnamefont {Christensen}}, \bibinfo {author} {\bibfnamefont
  {B.}~\bibnamefont {Grenier}}, \bibinfo {author} {\bibfnamefont {D.~F.}\
  \bibnamefont {McMorrow}},\ and\ \bibinfo {author} {\bibfnamefont
  {J.}~\bibnamefont {Mesot}},\ }\bibfield  {title} {\bibinfo {title}
  {Field-controlled magnetic order in the quantum spin-ladder system
  {Hpip}$_{2}${CuBr}$_{4}$},\ }\href
  {https://doi.org/10.1103/PhysRevB.79.020408} {\bibfield  {journal} {\bibinfo
  {journal} {Phys. Rev. B}\ }\textbf {\bibinfo {volume} {79}},\ \bibinfo
  {pages} {020408} (\bibinfo {year} {2009}{\natexlab{b}})}\BibitemShut
  {NoStop}%
\bibitem [{\citenamefont {Bewley}\ \emph {et~al.}(2011)\citenamefont {Bewley},
  \citenamefont {Taylor},\ and\ \citenamefont {Bennington}}]{bewley2011}%
  \BibitemOpen
  \bibfield  {author} {\bibinfo {author} {\bibfnamefont {R.~I.}\ \bibnamefont
  {Bewley}}, \bibinfo {author} {\bibfnamefont {J.~W.}\ \bibnamefont {Taylor}},\
  and\ \bibinfo {author} {\bibfnamefont {S.~M.}\ \bibnamefont {Bennington}},\
  }\bibfield  {title} {\bibinfo {title} {{LET}, a cold neutron multi-disk
  chopper spectrometer at {ISIS}},\ }\href
  {https://doi.org/http://dx.doi.org/10.1016/j.nima.2011.01.173} {\bibfield
  {journal} {\bibinfo  {journal} {Nucl. Instrum. Methods Phys. Res., Sect. A}\
  }\textbf {\bibinfo {volume} {637}},\ \bibinfo {pages} {128} (\bibinfo {year}
  {2011})}\BibitemShut {NoStop}%
\bibitem [{\citenamefont {Sutherland}(1975)}]{sutherland1975}%
  \BibitemOpen
  \bibfield  {author} {\bibinfo {author} {\bibfnamefont {B.}~\bibnamefont
  {Sutherland}},\ }\bibfield  {title} {\bibinfo {title} {{Model for a
  multicomponent quantum system}},\ }\href
  {https://doi.org/10.1103/PhysRevB.12.3795} {\bibfield  {journal} {\bibinfo
  {journal} {Phys. Rev. B}\ }\textbf {\bibinfo {volume} {12}},\ \bibinfo
  {pages} {3795} (\bibinfo {year} {1975})}\BibitemShut {NoStop}%
\bibitem [{\citenamefont {des Cloizeaux}\ and\ \citenamefont
  {Pearson}(1962)}]{dcp}%
  \BibitemOpen
  \bibfield  {author} {\bibinfo {author} {\bibfnamefont {J.}~\bibnamefont {des
  Cloizeaux}}\ and\ \bibinfo {author} {\bibfnamefont {J.~J.}\ \bibnamefont
  {Pearson}},\ }\bibfield  {title} {\bibinfo {title} {{Spin-Wave Spectrum of
  the Antiferromagnetic Linear Chain}},\ }\href
  {https://doi.org/10.1103/PhysRev.128.2131} {\bibfield  {journal} {\bibinfo
  {journal} {Phys. Rev.}\ }\textbf {\bibinfo {volume} {128}},\ \bibinfo {pages}
  {2131} (\bibinfo {year} {1962})}\BibitemShut {NoStop}%
\bibitem [{\citenamefont {White}\ and\ \citenamefont
  {Feiguin}(2004)}]{white2004}%
  \BibitemOpen
  \bibfield  {author} {\bibinfo {author} {\bibfnamefont {S.~R.}\ \bibnamefont
  {White}}\ and\ \bibinfo {author} {\bibfnamefont {A.~E.}\ \bibnamefont
  {Feiguin}},\ }\bibfield  {title} {\bibinfo {title} {{Real-Time Evolution
  Using the Density Matrix Renormalization Group}},\ }\href
  {https://doi.org/10.1103/PhysRevLett.93.076401} {\bibfield  {journal}
  {\bibinfo  {journal} {Phys. Rev. Lett.}\ }\textbf {\bibinfo {volume} {93}},\
  \bibinfo {pages} {076401} (\bibinfo {year} {2004})}\BibitemShut {NoStop}%
\bibitem [{\citenamefont {Daley}\ \emph {et~al.}(2004)\citenamefont {Daley},
  \citenamefont {Kollath}, \citenamefont {Schollw\"ock},\ and\ \citenamefont
  {Vidal}}]{daley2004}%
  \BibitemOpen
  \bibfield  {author} {\bibinfo {author} {\bibfnamefont {A.~J.}\ \bibnamefont
  {Daley}}, \bibinfo {author} {\bibfnamefont {C.}~\bibnamefont {Kollath}},
  \bibinfo {author} {\bibfnamefont {U.}~\bibnamefont {Schollw\"ock}},\ and\
  \bibinfo {author} {\bibfnamefont {G.}~\bibnamefont {Vidal}},\ }\bibfield
  {title} {\bibinfo {title} {Time-dependent density-matrix
  renormalization-group using adaptive effective {Hilbert} spaces},\ }\href
  {https://doi.org/10.1088/1742-5468/2004/04/P04005} {\bibfield  {journal}
  {\bibinfo  {journal} {J. Stat. Mech.: Theory Exp.}\ }\textbf {\bibinfo
  {volume} {2004}},\ \bibinfo {pages} {P04005}}\BibitemShut {NoStop}%
\bibitem [{\citenamefont {Schollw\"ock}(2011)}]{schollwock2011}%
  \BibitemOpen
  \bibfield  {author} {\bibinfo {author} {\bibfnamefont {U.}~\bibnamefont
  {Schollw\"ock}},\ }\bibfield  {title} {\bibinfo {title} {The density-matrix
  renormalization group in the age of matrix product states},\ }\href
  {https://doi.org/https://doi.org/10.1016/j.aop.2010.09.012} {\bibfield
  {journal} {\bibinfo  {journal} {Ann. Phys.}\ }\textbf {\bibinfo {volume}
  {326}},\ \bibinfo {pages} {96} (\bibinfo {year} {2011})}\BibitemShut
  {NoStop}%
\bibitem [{\citenamefont {Keselman}\ \emph {et~al.}(2020)\citenamefont
  {Keselman}, \citenamefont {Balents},\ and\ \citenamefont
  {Starykh}}]{keselman2020}%
  \BibitemOpen
  \bibfield  {author} {\bibinfo {author} {\bibfnamefont {A.}~\bibnamefont
  {Keselman}}, \bibinfo {author} {\bibfnamefont {L.}~\bibnamefont {Balents}},\
  and\ \bibinfo {author} {\bibfnamefont {O.~A.}\ \bibnamefont {Starykh}},\
  }\bibfield  {title} {\bibinfo {title} {{Dynamical Signatures of Quasiparticle
  Interactions in Quantum Spin Chains}},\ }\href
  {https://doi.org/10.1103/PhysRevLett.125.187201} {\bibfield  {journal}
  {\bibinfo  {journal} {Phys. Rev. Lett.}\ }\textbf {\bibinfo {volume} {125}},\
  \bibinfo {pages} {187201} (\bibinfo {year} {2020})}\BibitemShut {NoStop}%
\bibitem [{\citenamefont {Oshikawa}\ and\ \citenamefont
  {Affleck}(1997)}]{oshikawa1997}%
  \BibitemOpen
  \bibfield  {author} {\bibinfo {author} {\bibfnamefont {M.}~\bibnamefont
  {Oshikawa}}\ and\ \bibinfo {author} {\bibfnamefont {I.}~\bibnamefont
  {Affleck}},\ }\bibfield  {title} {\bibinfo {title} {{Field-Induced Gap in $S
  = 1/2$ Antiferromagnetic Chains}},\ }\href
  {https://doi.org/10.1103/PhysRevLett.79.2883} {\bibfield  {journal} {\bibinfo
   {journal} {Phys. Rev. Lett.}\ }\textbf {\bibinfo {volume} {79}},\ \bibinfo
  {pages} {2883} (\bibinfo {year} {1997})}\BibitemShut {NoStop}%
\bibitem [{\citenamefont {Sorella}\ and\ \citenamefont
  {Parola}(1998)}]{sorella1998}%
  \BibitemOpen
  \bibfield  {author} {\bibinfo {author} {\bibfnamefont {S.}~\bibnamefont
  {Sorella}}\ and\ \bibinfo {author} {\bibfnamefont {A.}~\bibnamefont
  {Parola}},\ }\bibfield  {title} {\bibinfo {title} {Theory of hole propagation
  in one-dimensional insulators and superconductors},\ }\href
  {https://doi.org/10.1103/PhysRevB.57.6444} {\bibfield  {journal} {\bibinfo
  {journal} {Phys. Rev. B}\ }\textbf {\bibinfo {volume} {57}},\ \bibinfo
  {pages} {6444} (\bibinfo {year} {1998})}\BibitemShut {NoStop}%
\bibitem [{\citenamefont {Saiga}\ and\ \citenamefont
  {Kuramoto}(1999)}]{saiga1999}%
  \BibitemOpen
  \bibfield  {author} {\bibinfo {author} {\bibfnamefont {Y.}~\bibnamefont
  {Saiga}}\ and\ \bibinfo {author} {\bibfnamefont {Y.}~\bibnamefont
  {Kuramoto}},\ }\bibfield  {title} {\bibinfo {title} {{Dynamical Properties of
  the One-Dimensional Supersymmetric $t$-$J$ Model: A View from Elementary
  Excitations}},\ }\href {https://doi.org/10.1143/JPSJ.68.3631} {\bibfield
  {journal} {\bibinfo  {journal} {J. Phys. Soc. Jpn.}\ }\textbf {\bibinfo
  {volume} {68}},\ \bibinfo {pages} {3631} (\bibinfo {year}
  {1999})}\BibitemShut {NoStop}%
\bibitem [{\citenamefont {Barthel}(2013)}]{barthel2013}%
  \BibitemOpen
  \bibfield  {author} {\bibinfo {author} {\bibfnamefont {T.}~\bibnamefont
  {Barthel}},\ }\bibfield  {title} {\bibinfo {title} {Precise evaluation of
  thermal response functions by optimized density matrix renormalization group
  schemes},\ }\href@noop {} {\bibfield  {journal} {\bibinfo  {journal} {New J.
  Phys.}\ }\textbf {\bibinfo {volume} {15}},\ \bibinfo {pages} {073010}
  (\bibinfo {year} {2013})}\BibitemShut {NoStop}%
\bibitem [{\citenamefont {Kestin}\ and\ \citenamefont
  {Giamarchi}(2019)}]{kestin2019}%
  \BibitemOpen
  \bibfield  {author} {\bibinfo {author} {\bibfnamefont {N.}~\bibnamefont
  {Kestin}}\ and\ \bibinfo {author} {\bibfnamefont {T.}~\bibnamefont
  {Giamarchi}},\ }\bibfield  {title} {\bibinfo {title} {Low-dimensional
  correlations under thermal fluctuations},\ }\href
  {https://doi.org/10.1103/PhysRevB.99.195121} {\bibfield  {journal} {\bibinfo
  {journal} {Phys. Rev. B}\ }\textbf {\bibinfo {volume} {99}},\ \bibinfo
  {pages} {195121} (\bibinfo {year} {2019})}\BibitemShut {NoStop}%
\bibitem [{\citenamefont {Verstraete}\ \emph {et~al.}(2004)\citenamefont
  {Verstraete}, \citenamefont {Garcia-Ripoll},\ and\ \citenamefont
  {Cirac}}]{verstraete2004}%
  \BibitemOpen
  \bibfield  {author} {\bibinfo {author} {\bibfnamefont {F.}~\bibnamefont
  {Verstraete}}, \bibinfo {author} {\bibfnamefont {J.~J.}\ \bibnamefont
  {Garcia-Ripoll}},\ and\ \bibinfo {author} {\bibfnamefont {J.~I.}\
  \bibnamefont {Cirac}},\ }\bibfield  {title} {\bibinfo {title} {Matrix
  {Product} {Density} {Operators}: {Simulation} of {Finite}-{Temperature} and
  {Dissipative} {Systems}},\ }\href
  {https://doi.org/10.1103/PhysRevLett.93.207204} {\bibfield  {journal}
  {\bibinfo  {journal} {Phys. Rev. Lett.}\ }\textbf {\bibinfo {volume} {93}},\
  \bibinfo {pages} {207204} (\bibinfo {year} {2004})}\BibitemShut {NoStop}%
\bibitem [{\citenamefont {Zwolak}\ and\ \citenamefont
  {Vidal}(2004)}]{zwolak2004}%
  \BibitemOpen
  \bibfield  {author} {\bibinfo {author} {\bibfnamefont {M.}~\bibnamefont
  {Zwolak}}\ and\ \bibinfo {author} {\bibfnamefont {G.}~\bibnamefont {Vidal}},\
  }\bibfield  {title} {\bibinfo {title} {Mixed-state dynamics in
  one-dimensional quantum lattice systems: A time-dependent superoperator
  renormalization algorithm},\ }\href
  {https://doi.org/10.1103/PhysRevLett.93.207205} {\bibfield  {journal}
  {\bibinfo  {journal} {Phys. Rev. Lett.}\ }\textbf {\bibinfo {volume} {93}},\
  \bibinfo {pages} {207205} (\bibinfo {year} {2004})}\BibitemShut {NoStop}%
\bibitem [{\citenamefont {Feiguin}\ and\ \citenamefont
  {Fiete}(2010)}]{feiguin2010}%
  \BibitemOpen
  \bibfield  {author} {\bibinfo {author} {\bibfnamefont {A.~E.}\ \bibnamefont
  {Feiguin}}\ and\ \bibinfo {author} {\bibfnamefont {G.}~\bibnamefont
  {Fiete}},\ }\bibfield  {title} {\bibinfo {title} {{Spectral properties of a
  spin-incoherent Luttinger liquid}},\ }\href
  {https://doi.org/10.1103/PhysRevB.81.075108} {\bibfield  {journal} {\bibinfo
  {journal} {Phys. Rev. B}\ }\textbf {\bibinfo {volume} {81}},\ \bibinfo
  {pages} {075108} (\bibinfo {year} {2010})}\BibitemShut {NoStop}%
\bibitem [{\citenamefont {Cheianov}\ and\ \citenamefont
  {Zvonarev}(2004)}]{cheianov2004}%
  \BibitemOpen
  \bibfield  {author} {\bibinfo {author} {\bibfnamefont {V.~V.}\ \bibnamefont
  {Cheianov}}\ and\ \bibinfo {author} {\bibfnamefont {M.~B.}\ \bibnamefont
  {Zvonarev}},\ }\bibfield  {title} {\bibinfo {title} {{Nonunitary Spin-Charge
  Separation in a One-Dimensional Fermion Gas}},\ }\href
  {https://doi.org/10.1103/PhysRevLett.92.176401} {\bibfield  {journal}
  {\bibinfo  {journal} {Phys. Rev. Lett.}\ }\textbf {\bibinfo {volume} {92}},\
  \bibinfo {pages} {176401} (\bibinfo {year} {2004})}\BibitemShut {NoStop}%
\bibitem [{\citenamefont {Fiete}(2007)}]{fiete2007}%
  \BibitemOpen
  \bibfield  {author} {\bibinfo {author} {\bibfnamefont {G.}~\bibnamefont
  {Fiete}},\ }\bibfield  {title} {\bibinfo {title} {{The spin-incoherent
  Luttinger liquid}},\ }\href {https://doi.org/10.1103/RevModPhys.79.801}
  {\bibfield  {journal} {\bibinfo  {journal} {Rev. Mod. Phys.}\ }\textbf
  {\bibinfo {volume} {79}},\ \bibinfo {pages} {801} (\bibinfo {year}
  {2007})}\BibitemShut {NoStop}%
\bibitem [{\citenamefont {Patyal}\ \emph {et~al.}(1990)\citenamefont {Patyal},
  \citenamefont {Scott},\ and\ \citenamefont {Willett}}]{patyal1990}%
  \BibitemOpen
  \bibfield  {author} {\bibinfo {author} {\bibfnamefont {B.~R.}\ \bibnamefont
  {Patyal}}, \bibinfo {author} {\bibfnamefont {B.~L.}\ \bibnamefont {Scott}},\
  and\ \bibinfo {author} {\bibfnamefont {R.~D.}\ \bibnamefont {Willett}},\
  }\bibfield  {title} {\bibinfo {title} {Crystal-structure,
  magnetic-susceptibility, and {EPR} studies of
  bis(piperidinium)tetrabromocuprate({II}): {A} novel monomer system showing
  spin diffusion},\ }\href {https://doi.org/10.1103/PhysRevB.41.1657}
  {\bibfield  {journal} {\bibinfo  {journal} {Phys. Rev. B}\ }\textbf {\bibinfo
  {volume} {41}},\ \bibinfo {pages} {1657} (\bibinfo {year}
  {1990})}\BibitemShut {NoStop}%
\bibitem [{\citenamefont {Arnold}\ \emph {et~al.}(2014)\citenamefont {Arnold},
  \citenamefont {Bilheux}, \citenamefont {Borreguero}, \citenamefont {Buts},
  \citenamefont {Campbell}, \citenamefont {Chapon}, \citenamefont {Doucet},
  \citenamefont {Draper}, \citenamefont {Leal}, \citenamefont {Gigg},
  \citenamefont {Lynch}, \citenamefont {Markvardsen}, \citenamefont
  {Mikkelson}, \citenamefont {Mikkelson}, \citenamefont {Miller}, \citenamefont
  {Palmen}, \citenamefont {Parker}, \citenamefont {Passos}, \citenamefont
  {Perring}, \citenamefont {Peterson}, \citenamefont {Ren}, \citenamefont
  {Reuter}, \citenamefont {Savici}, \citenamefont {Taylor}, \citenamefont
  {Taylor}, \citenamefont {Tolchenov}, \citenamefont {Zhou},\ and\
  \citenamefont {Zikovsky}}]{arnold2014}%
  \BibitemOpen
  \bibfield  {author} {\bibinfo {author} {\bibfnamefont {O.}~\bibnamefont
  {Arnold}}, \bibinfo {author} {\bibfnamefont {J.~C.}\ \bibnamefont {Bilheux}},
  \bibinfo {author} {\bibfnamefont {J.~M.}\ \bibnamefont {Borreguero}},
  \bibinfo {author} {\bibfnamefont {A.}~\bibnamefont {Buts}}, \bibinfo {author}
  {\bibfnamefont {S.~I.}\ \bibnamefont {Campbell}}, \bibinfo {author}
  {\bibfnamefont {L.}~\bibnamefont {Chapon}}, \bibinfo {author} {\bibfnamefont
  {M.}~\bibnamefont {Doucet}}, \bibinfo {author} {\bibfnamefont
  {N.}~\bibnamefont {Draper}}, \bibinfo {author} {\bibfnamefont {R.~F.}\
  \bibnamefont {Leal}}, \bibinfo {author} {\bibfnamefont {M.~A.}\ \bibnamefont
  {Gigg}}, \bibinfo {author} {\bibfnamefont {V.~E.}\ \bibnamefont {Lynch}},
  \bibinfo {author} {\bibfnamefont {A.}~\bibnamefont {Markvardsen}}, \bibinfo
  {author} {\bibfnamefont {D.~J.}\ \bibnamefont {Mikkelson}}, \bibinfo {author}
  {\bibfnamefont {R.~L.}\ \bibnamefont {Mikkelson}}, \bibinfo {author}
  {\bibfnamefont {R.}~\bibnamefont {Miller}}, \bibinfo {author} {\bibfnamefont
  {K.}~\bibnamefont {Palmen}}, \bibinfo {author} {\bibfnamefont
  {P.}~\bibnamefont {Parker}}, \bibinfo {author} {\bibfnamefont
  {G.}~\bibnamefont {Passos}}, \bibinfo {author} {\bibfnamefont {T.~G.}\
  \bibnamefont {Perring}}, \bibinfo {author} {\bibfnamefont {P.~F.}\
  \bibnamefont {Peterson}}, \bibinfo {author} {\bibfnamefont {S.}~\bibnamefont
  {Ren}}, \bibinfo {author} {\bibfnamefont {M.~A.}\ \bibnamefont {Reuter}},
  \bibinfo {author} {\bibfnamefont {A.~T.}\ \bibnamefont {Savici}}, \bibinfo
  {author} {\bibfnamefont {J.~W.}\ \bibnamefont {Taylor}}, \bibinfo {author}
  {\bibfnamefont {R.~J.}\ \bibnamefont {Taylor}}, \bibinfo {author}
  {\bibfnamefont {R.}~\bibnamefont {Tolchenov}}, \bibinfo {author}
  {\bibfnamefont {W.}~\bibnamefont {Zhou}},\ and\ \bibinfo {author}
  {\bibfnamefont {J.}~\bibnamefont {Zikovsky}},\ }\bibfield  {title} {\bibinfo
  {title} {Mantid -- data analysis and visualization package for neutron
  scattering and $\mu${SR} experiments},\ }\href
  {https://doi.org/https://doi.org/10.1016/j.nima.2014.07.029} {\bibfield
  {journal} {\bibinfo  {journal} {Nucl. Instrum. Methods Phys. Res., Sect. A}\
  }\textbf {\bibinfo {volume} {764}},\ \bibinfo {pages} {156} (\bibinfo {year}
  {2014})}\BibitemShut {NoStop}%
\bibitem [{\citenamefont {Ewings}\ \emph {et~al.}(2016)\citenamefont {Ewings},
  \citenamefont {Buts}, \citenamefont {Le}, \citenamefont {Duijn},
  \citenamefont {Bustinduy},\ and\ \citenamefont {Perring}}]{ewings2016}%
  \BibitemOpen
  \bibfield  {author} {\bibinfo {author} {\bibfnamefont {R.~A.}\ \bibnamefont
  {Ewings}}, \bibinfo {author} {\bibfnamefont {A.}~\bibnamefont {Buts}},
  \bibinfo {author} {\bibfnamefont {M.~D.}\ \bibnamefont {Le}}, \bibinfo
  {author} {\bibfnamefont {J.~v.}\ \bibnamefont {Duijn}}, \bibinfo {author}
  {\bibfnamefont {I.}~\bibnamefont {Bustinduy}},\ and\ \bibinfo {author}
  {\bibfnamefont {T.~G.}\ \bibnamefont {Perring}},\ }\bibfield  {title}
  {\bibinfo {title} {Horace: {Software} for the analysis of data from single
  crystal spectroscopy experiments at time-of-flight neutron instruments},\
  }\href {https://doi.org/http://dx.doi.org/10.1016/j.nima.2016.07.036}
  {\bibfield  {journal} {\bibinfo  {journal} {Nucl. Instrum. Methods Phys.
  Res., Sect. A}\ }\textbf {\bibinfo {volume} {834}},\ \bibinfo {pages} {132}
  (\bibinfo {year} {2016})}\BibitemShut {NoStop}%
\bibitem [{\citenamefont {Fishman}\ \emph {et~al.}(2022)\citenamefont
  {Fishman}, \citenamefont {White},\ and\ \citenamefont
  {Stoudenmire}}]{itensor}%
  \BibitemOpen
  \bibfield  {author} {\bibinfo {author} {\bibfnamefont {M.}~\bibnamefont
  {Fishman}}, \bibinfo {author} {\bibfnamefont {S.~R.}\ \bibnamefont {White}},\
  and\ \bibinfo {author} {\bibfnamefont {E.~M.}\ \bibnamefont {Stoudenmire}},\
  }\bibfield  {title} {\bibinfo {title} {{The ITensor Software Library for
  Tensor Network Calculations}},\ }\href
  {https://doi.org/10.21468/SciPostPhysCodeb.4} {\bibfield  {journal} {\bibinfo
   {journal} {SciPost Phys. Codebases}\ ,\ \bibinfo {pages} {4}} (\bibinfo
  {year} {2022})}\BibitemShut {NoStop}%
\bibitem [{\citenamefont {White}\ and\ \citenamefont
  {Affleck}(2008)}]{white2008}%
  \BibitemOpen
  \bibfield  {author} {\bibinfo {author} {\bibfnamefont {S.~R.}\ \bibnamefont
  {White}}\ and\ \bibinfo {author} {\bibfnamefont {I.}~\bibnamefont
  {Affleck}},\ }\bibfield  {title} {\bibinfo {title} {{Spectral function for
  the $S = 1$ Heisenberg antiferromagetic chain}},\ }\href
  {https://doi.org/10.1103/PhysRevB.77.134437} {\bibfield  {journal} {\bibinfo
  {journal} {Phys. Rev. B}\ }\textbf {\bibinfo {volume} {77}},\ \bibinfo
  {pages} {134437} (\bibinfo {year} {2008})}\BibitemShut {NoStop}%
\end{thebibliography}%


\begin{thebibliography}{33}%
\makeatletter
\providecommand \@ifxundefined [1]{%
 \@ifx{#1\undefined}
}%
\providecommand \@ifnum [1]{%
 \ifnum #1\expandafter \@firstoftwo
 \else \expandafter \@secondoftwo
 \fi
}%
\providecommand \@ifx [1]{%
 \ifx #1\expandafter \@firstoftwo
 \else \expandafter \@secondoftwo
 \fi
}%
\providecommand \natexlab [1]{#1}%
\providecommand \enquote  [1]{``#1''}%
\providecommand \bibnamefont  [1]{#1}%
\providecommand \bibfnamefont [1]{#1}%
\providecommand \citenamefont [1]{#1}%
\providecommand \href@noop [0]{\@secondoftwo}%
\providecommand \href [0]{\begingroup \@sanitize@url \@href}%
\providecommand \@href[1]{\@@startlink{#1}\@@href}%
\providecommand \@@href[1]{\endgroup#1\@@endlink}%
\providecommand \@sanitize@url [0]{\catcode `\\12\catcode `\$12\catcode
  `\&12\catcode `\#12\catcode `\^12\catcode `\_12\catcode `\%12\relax}%
\providecommand \@@startlink[1]{}%
\providecommand \@@endlink[0]{}%
\providecommand \url  [0]{\begingroup\@sanitize@url \@url }%
\providecommand \@url [1]{\endgroup\@href {#1}{\urlprefix }}%
\providecommand \urlprefix  [0]{URL }%
\providecommand \Eprint [0]{\href }%
\providecommand \doibase [0]{https://doi.org/}%
\providecommand \selectlanguage [0]{\@gobble}%
\providecommand \bibinfo  [0]{\@secondoftwo}%
\providecommand \bibfield  [0]{\@secondoftwo}%
\providecommand \translation [1]{[#1]}%
\providecommand \BibitemOpen [0]{}%
\providecommand \bibitemStop [0]{}%
\providecommand \bibitemNoStop [0]{.\EOS\space}%
\providecommand \EOS [0]{\spacefactor3000\relax}%
\providecommand \BibitemShut  [1]{\csname bibitem#1\endcsname}%
\let\auto@bib@innerbib\@empty
\bibitem [{\citenamefont {Patyal}\ \emph {et~al.}(1990)\citenamefont {Patyal},
  \citenamefont {Scott},\ and\ \citenamefont {Willett}}]{patyal1990}%
  \BibitemOpen
  \bibfield  {author} {\bibinfo {author} {\bibfnamefont {B.~R.}\ \bibnamefont
  {Patyal}}, \bibinfo {author} {\bibfnamefont {B.~L.}\ \bibnamefont {Scott}},\
  and\ \bibinfo {author} {\bibfnamefont {R.~D.}\ \bibnamefont {Willett}},\
  }\bibfield  {title} {\bibinfo {title} {Crystal-structure,
  magnetic-susceptibility, and {EPR} studies of
  bis(piperidinium)tetrabromocuprate({II}): {A} novel monomer system showing
  spin diffusion},\ }\href {https://doi.org/10.1103/PhysRevB.41.1657}
  {\bibfield  {journal} {\bibinfo  {journal} {Phys. Rev. B}\ }\textbf {\bibinfo
  {volume} {41}},\ \bibinfo {pages} {1657} (\bibinfo {year}
  {1990})}\BibitemShut {NoStop}%
\bibitem [{\citenamefont {Ward}\ \emph {et~al.}(2013)\citenamefont {Ward},
  \citenamefont {Bouillot}, \citenamefont {Ryll}, \citenamefont {Kiefer},
  \citenamefont {Kr\"amer}, \citenamefont {R\"uegg}, \citenamefont {Kollath},\
  and\ \citenamefont {Giamarchi}}]{ward2013}%
  \BibitemOpen
  \bibfield  {author} {\bibinfo {author} {\bibfnamefont {S.}~\bibnamefont
  {Ward}}, \bibinfo {author} {\bibfnamefont {P.}~\bibnamefont {Bouillot}},
  \bibinfo {author} {\bibfnamefont {H.}~\bibnamefont {Ryll}}, \bibinfo {author}
  {\bibfnamefont {K.}~\bibnamefont {Kiefer}}, \bibinfo {author} {\bibfnamefont
  {K.~W.}\ \bibnamefont {Kr\"amer}}, \bibinfo {author} {\bibfnamefont
  {C.}~\bibnamefont {R\"uegg}}, \bibinfo {author} {\bibfnamefont
  {C.}~\bibnamefont {Kollath}},\ and\ \bibinfo {author} {\bibfnamefont
  {T.}~\bibnamefont {Giamarchi}},\ }\bibfield  {title} {\bibinfo {title} {Spin
  ladders and quantum simulators for {Tomonaga}-{Luttinger} liquids},\ }\href
  {https://doi.org/10.1088/0953-8984/25/1/014004} {\bibfield  {journal}
  {\bibinfo  {journal} {J. Phys.: Condens. Matter}\ }\textbf {\bibinfo {volume}
  {25}},\ \bibinfo {pages} {014004} (\bibinfo {year} {2013})}\BibitemShut
  {NoStop}%
\bibitem [{\citenamefont {Ward}\ \emph {et~al.}(2017)\citenamefont {Ward},
  \citenamefont {Mena}, \citenamefont {Bouillot}, \citenamefont {Kollath},
  \citenamefont {Giamarchi}, \citenamefont {Schmidt}, \citenamefont {Normand},
  \citenamefont {Kr\"amer}, \citenamefont {Biner}, \citenamefont {Bewley},
  \citenamefont {Guidi}, \citenamefont {Boehm}, \citenamefont {McMorrow},\ and\
  \citenamefont {R\"uegg}}]{ward2017}%
  \BibitemOpen
  \bibfield  {author} {\bibinfo {author} {\bibfnamefont {S.}~\bibnamefont
  {Ward}}, \bibinfo {author} {\bibfnamefont {M.}~\bibnamefont {Mena}}, \bibinfo
  {author} {\bibfnamefont {P.}~\bibnamefont {Bouillot}}, \bibinfo {author}
  {\bibfnamefont {C.}~\bibnamefont {Kollath}}, \bibinfo {author} {\bibfnamefont
  {T.}~\bibnamefont {Giamarchi}}, \bibinfo {author} {\bibfnamefont {K.~P.}\
  \bibnamefont {Schmidt}}, \bibinfo {author} {\bibfnamefont {B.}~\bibnamefont
  {Normand}}, \bibinfo {author} {\bibfnamefont {K.~W.}\ \bibnamefont
  {Kr\"amer}}, \bibinfo {author} {\bibfnamefont {D.}~\bibnamefont {Biner}},
  \bibinfo {author} {\bibfnamefont {R.}~\bibnamefont {Bewley}}, \bibinfo
  {author} {\bibfnamefont {T.}~\bibnamefont {Guidi}}, \bibinfo {author}
  {\bibfnamefont {M.}~\bibnamefont {Boehm}}, \bibinfo {author} {\bibfnamefont
  {D.~F.}\ \bibnamefont {McMorrow}},\ and\ \bibinfo {author} {\bibfnamefont
  {C.}~\bibnamefont {R\"uegg}},\ }\bibfield  {title} {\bibinfo {title} {Bound
  {States} and {Field}-{Polarized} {Haldane} {Modes} in a {Quantum} {Spin}
  {Ladder}},\ }\href {https://doi.org/10.1103/PhysRevLett.118.177202}
  {\bibfield  {journal} {\bibinfo  {journal} {Phys. Rev. Lett.}\ }\textbf
  {\bibinfo {volume} {118}},\ \bibinfo {pages} {177202} (\bibinfo {year}
  {2017})}\BibitemShut {NoStop}%
\bibitem [{\citenamefont {Bewley}\ \emph {et~al.}(2011)\citenamefont {Bewley},
  \citenamefont {Taylor},\ and\ \citenamefont {Bennington}}]{bewley2011}%
  \BibitemOpen
  \bibfield  {author} {\bibinfo {author} {\bibfnamefont {R.~I.}\ \bibnamefont
  {Bewley}}, \bibinfo {author} {\bibfnamefont {J.~W.}\ \bibnamefont {Taylor}},\
  and\ \bibinfo {author} {\bibfnamefont {S.~M.}\ \bibnamefont {Bennington}},\
  }\bibfield  {title} {\bibinfo {title} {{LET}, a cold neutron multi-disk
  chopper spectrometer at {ISIS}},\ }\href
  {https://doi.org/http://dx.doi.org/10.1016/j.nima.2011.01.173} {\bibfield
  {journal} {\bibinfo  {journal} {Nucl. Instrum. Methods Phys. Res., Sect. A}\
  }\textbf {\bibinfo {volume} {637}},\ \bibinfo {pages} {128} (\bibinfo {year}
  {2011})}\BibitemShut {NoStop}%
\bibitem [{\citenamefont {Arnold}\ \emph {et~al.}(2014)\citenamefont {Arnold},
  \citenamefont {Bilheux}, \citenamefont {Borreguero}, \citenamefont {Buts},
  \citenamefont {Campbell}, \citenamefont {Chapon}, \citenamefont {Doucet},
  \citenamefont {Draper}, \citenamefont {Leal}, \citenamefont {Gigg},
  \citenamefont {Lynch}, \citenamefont {Markvardsen}, \citenamefont
  {Mikkelson}, \citenamefont {Mikkelson}, \citenamefont {Miller}, \citenamefont
  {Palmen}, \citenamefont {Parker}, \citenamefont {Passos}, \citenamefont
  {Perring}, \citenamefont {Peterson}, \citenamefont {Ren}, \citenamefont
  {Reuter}, \citenamefont {Savici}, \citenamefont {Taylor}, \citenamefont
  {Taylor}, \citenamefont {Tolchenov}, \citenamefont {Zhou},\ and\
  \citenamefont {Zikovsky}}]{arnold2014}%
  \BibitemOpen
  \bibfield  {author} {\bibinfo {author} {\bibfnamefont {O.}~\bibnamefont
  {Arnold}}, \bibinfo {author} {\bibfnamefont {J.~C.}\ \bibnamefont {Bilheux}},
  \bibinfo {author} {\bibfnamefont {J.~M.}\ \bibnamefont {Borreguero}},
  \bibinfo {author} {\bibfnamefont {A.}~\bibnamefont {Buts}}, \bibinfo {author}
  {\bibfnamefont {S.~I.}\ \bibnamefont {Campbell}}, \bibinfo {author}
  {\bibfnamefont {L.}~\bibnamefont {Chapon}}, \bibinfo {author} {\bibfnamefont
  {M.}~\bibnamefont {Doucet}}, \bibinfo {author} {\bibfnamefont
  {N.}~\bibnamefont {Draper}}, \bibinfo {author} {\bibfnamefont {R.~F.}\
  \bibnamefont {Leal}}, \bibinfo {author} {\bibfnamefont {M.~A.}\ \bibnamefont
  {Gigg}}, \bibinfo {author} {\bibfnamefont {V.~E.}\ \bibnamefont {Lynch}},
  \bibinfo {author} {\bibfnamefont {A.}~\bibnamefont {Markvardsen}}, \bibinfo
  {author} {\bibfnamefont {D.~J.}\ \bibnamefont {Mikkelson}}, \bibinfo {author}
  {\bibfnamefont {R.~L.}\ \bibnamefont {Mikkelson}}, \bibinfo {author}
  {\bibfnamefont {R.}~\bibnamefont {Miller}}, \bibinfo {author} {\bibfnamefont
  {K.}~\bibnamefont {Palmen}}, \bibinfo {author} {\bibfnamefont
  {P.}~\bibnamefont {Parker}}, \bibinfo {author} {\bibfnamefont
  {G.}~\bibnamefont {Passos}}, \bibinfo {author} {\bibfnamefont {T.~G.}\
  \bibnamefont {Perring}}, \bibinfo {author} {\bibfnamefont {P.~F.}\
  \bibnamefont {Peterson}}, \bibinfo {author} {\bibfnamefont {S.}~\bibnamefont
  {Ren}}, \bibinfo {author} {\bibfnamefont {M.~A.}\ \bibnamefont {Reuter}},
  \bibinfo {author} {\bibfnamefont {A.~T.}\ \bibnamefont {Savici}}, \bibinfo
  {author} {\bibfnamefont {J.~W.}\ \bibnamefont {Taylor}}, \bibinfo {author}
  {\bibfnamefont {R.~J.}\ \bibnamefont {Taylor}}, \bibinfo {author}
  {\bibfnamefont {R.}~\bibnamefont {Tolchenov}}, \bibinfo {author}
  {\bibfnamefont {W.}~\bibnamefont {Zhou}},\ and\ \bibinfo {author}
  {\bibfnamefont {J.}~\bibnamefont {Zikovsky}},\ }\bibfield  {title} {\bibinfo
  {title} {Mantid -- data analysis and visualization package for neutron
  scattering and $\mu${SR} experiments},\ }\href
  {https://doi.org/https://doi.org/10.1016/j.nima.2014.07.029} {\bibfield
  {journal} {\bibinfo  {journal} {Nucl. Instrum. Methods Phys. Res., Sect. A}\
  }\textbf {\bibinfo {volume} {764}},\ \bibinfo {pages} {156} (\bibinfo {year}
  {2014})}\BibitemShut {NoStop}%
\bibitem [{Ton()}]{Tong}%
  \BibitemOpen
  \bibinfo {title} {{D. Tong, Lectures on Supersymmetric Quantum Mechanics,
  https://www.damtp.cam.ac.uk/user/tong/ susyqm.html}}\BibitemShut {NoStop}%
\bibitem [{\citenamefont {Cai}\ \emph {et~al.}(2022)\citenamefont {Cai},
  \citenamefont {Wu}, \citenamefont {Mei}, \citenamefont {Zhao}, \citenamefont
  {Jiang}, \citenamefont {Yao}, \citenamefont {He}, \citenamefont {Zhou},\ and\
  \citenamefont {Duan}}]{Cai2022}%
  \BibitemOpen
\bibfield  {title} {  }\bibfield  {author} {\bibinfo {author} {\bibfnamefont
  {M.-L.}\ \bibnamefont {Cai}}, \bibinfo {author} {\bibfnamefont {Y.-K.}\
  \bibnamefont {Wu}}, \bibinfo {author} {\bibfnamefont {Q.-X.}\ \bibnamefont
  {Mei}}, \bibinfo {author} {\bibfnamefont {W.-D.}\ \bibnamefont {Zhao}},
  \bibinfo {author} {\bibfnamefont {Y.}~\bibnamefont {Jiang}}, \bibinfo
  {author} {\bibfnamefont {L.}~\bibnamefont {Yao}}, \bibinfo {author}
  {\bibfnamefont {L.}~\bibnamefont {He}}, \bibinfo {author} {\bibfnamefont
  {Z.-C.}\ \bibnamefont {Zhou}},\ and\ \bibinfo {author} {\bibfnamefont
  {L.-M.}\ \bibnamefont {Duan}},\ }\bibfield  {title} {\bibinfo {title}
  {Observation of supersymmetry and its spontaneous breaking in a trapped ion
  quantum simulator},\ }\href {https://doi.org/10.1038/s41467-022-31058-0}
  {\bibfield  {journal} {\bibinfo  {journal} {Nature Commun.}\ }\textbf
  {\bibinfo {volume} {13}},\ \bibinfo {pages} {3412} (\bibinfo {year}
  {2022})}\BibitemShut {NoStop}%
\bibitem [{\citenamefont {Bares}\ and\ \citenamefont
  {Blatter}(1990)}]{bares1990}%
  \BibitemOpen
  \bibfield  {author} {\bibinfo {author} {\bibfnamefont {P.~A.}\ \bibnamefont
  {Bares}}\ and\ \bibinfo {author} {\bibfnamefont {G.}~\bibnamefont
  {Blatter}},\ }\bibfield  {title} {\bibinfo {title} {{Supersymmetric $t$-$J$
  model in one dimension: Separation of spin and charge}},\ }\href
  {https://doi.org/10.1103/PhysRevLett.64.2567} {\bibfield  {journal} {\bibinfo
   {journal} {Phys. Rev. Lett.}\ }\textbf {\bibinfo {volume} {64}},\ \bibinfo
  {pages} {2567} (\bibinfo {year} {1990})}\BibitemShut {NoStop}%
\bibitem [{\citenamefont {Bares}\ \emph {et~al.}(1991)\citenamefont {Bares},
  \citenamefont {Blatter},\ and\ \citenamefont {Ogata}}]{bares1991}%
  \BibitemOpen
  \bibfield  {author} {\bibinfo {author} {\bibfnamefont {P.-A.}\ \bibnamefont
  {Bares}}, \bibinfo {author} {\bibfnamefont {G.}~\bibnamefont {Blatter}},\
  and\ \bibinfo {author} {\bibfnamefont {M.}~\bibnamefont {Ogata}},\ }\bibfield
   {title} {\bibinfo {title} {{Exact solution of the $t$-$J$ model in one
  dimension at $2t = \pm J$: Ground state and excitation spectrum}},\ }\href
  {https://doi.org/10.1103/PhysRevB.44.130} {\bibfield  {journal} {\bibinfo
  {journal} {Phys. Rev. B}\ }\textbf {\bibinfo {volume} {44}},\ \bibinfo
  {pages} {130} (\bibinfo {year} {1991})}\BibitemShut {NoStop}%
\bibitem [{\citenamefont {Essler}\ and\ \citenamefont
  {Korepin}(1992)}]{essler1992}%
  \BibitemOpen
  \bibfield  {author} {\bibinfo {author} {\bibfnamefont {F.~H.~L.}\
  \bibnamefont {Essler}}\ and\ \bibinfo {author} {\bibfnamefont {V.~E.}\
  \bibnamefont {Korepin}},\ }\bibfield  {title} {\bibinfo {title} {{Higher
  conservation laws and algebraic Bethe Ans\"atze for the supersymmetric
  $t$-$J$ model}},\ }\href {https://doi.org/10.1103/PhysRevB.46.9147}
  {\bibfield  {journal} {\bibinfo  {journal} {Phys. Rev. B}\ }\textbf {\bibinfo
  {volume} {46}},\ \bibinfo {pages} {9147} (\bibinfo {year}
  {1992})}\BibitemShut {NoStop}%
\bibitem [{\citenamefont {Bouillot}\ \emph {et~al.}(2011)\citenamefont
  {Bouillot}, \citenamefont {Kollath}, \citenamefont {L\"auchli}, \citenamefont
  {Zvonarev}, \citenamefont {Thielemann}, \citenamefont {R\"uegg},
  \citenamefont {Orignac}, \citenamefont {Citro}, \citenamefont
  {Klanj\ifmmode~\check{s}\else \v{s}\fi{}ek}, \citenamefont {Berthier},
  \citenamefont {Horvati\ifmmode~\acute{c}\else \'{c}\fi{}},\ and\
  \citenamefont {Giamarchi}}]{bouillot2011}%
  \BibitemOpen
  \bibfield  {author} {\bibinfo {author} {\bibfnamefont {P.}~\bibnamefont
  {Bouillot}}, \bibinfo {author} {\bibfnamefont {C.}~\bibnamefont {Kollath}},
  \bibinfo {author} {\bibfnamefont {A.~M.}\ \bibnamefont {L\"auchli}}, \bibinfo
  {author} {\bibfnamefont {M.}~\bibnamefont {Zvonarev}}, \bibinfo {author}
  {\bibfnamefont {B.}~\bibnamefont {Thielemann}}, \bibinfo {author}
  {\bibfnamefont {C.}~\bibnamefont {R\"uegg}}, \bibinfo {author} {\bibfnamefont
  {E.}~\bibnamefont {Orignac}}, \bibinfo {author} {\bibfnamefont
  {R.}~\bibnamefont {Citro}}, \bibinfo {author} {\bibfnamefont
  {M.}~\bibnamefont {Klanj\ifmmode~\check{s}\else \v{s}\fi{}ek}}, \bibinfo
  {author} {\bibfnamefont {C.}~\bibnamefont {Berthier}}, \bibinfo {author}
  {\bibfnamefont {M.}~\bibnamefont {Horvati\ifmmode~\acute{c}\else
  \'{c}\fi{}}},\ and\ \bibinfo {author} {\bibfnamefont {T.}~\bibnamefont
  {Giamarchi}},\ }\bibfield  {title} {\bibinfo {title} {Statics and dynamics of
  weakly coupled antiferromagnetic spin-$\frac{1}{2}$ ladders in a magnetic
  field},\ }\href {https://doi.org/10.1103/PhysRevB.83.054407} {\bibfield
  {journal} {\bibinfo  {journal} {Phys. Rev. B}\ }\textbf {\bibinfo {volume}
  {83}},\ \bibinfo {pages} {054407} (\bibinfo {year} {2011})}\BibitemShut
  {NoStop}%
\bibitem [{\citenamefont {Batchelor}\ \emph {et~al.}(2007)\citenamefont
  {Batchelor}, \citenamefont {Guan}, \citenamefont {Oelkers},\ and\
  \citenamefont {Tsuboi}}]{Batchelor2007}%
  \BibitemOpen
  \bibfield  {author} {\bibinfo {author} {\bibfnamefont {M.~T.}\ \bibnamefont
  {Batchelor}}, \bibinfo {author} {\bibfnamefont {X.-W.}\ \bibnamefont {Guan}},
  \bibinfo {author} {\bibfnamefont {N.}~\bibnamefont {Oelkers}},\ and\ \bibinfo
  {author} {\bibfnamefont {Z.}~\bibnamefont {Tsuboi}},\ }\bibfield  {title}
  {\bibinfo {title} {{Integrable models and quantum spin ladders: comparison
  between theory and experiment for the strong coupling ladder compounds}},\
  }\href {https://doi.org/10.1080/00018730701265383} {\bibfield  {journal}
  {\bibinfo  {journal} {Adv. Phys.}\ }\textbf {\bibinfo {volume} {56}},\
  \bibinfo {pages} {465} (\bibinfo {year} {2007})}\BibitemShut {NoStop}%
\bibitem [{\citenamefont {des Cloizeaux}\ and\ \citenamefont
  {Pearson}(1962)}]{dcp}%
  \BibitemOpen
  \bibfield  {author} {\bibinfo {author} {\bibfnamefont {J.}~\bibnamefont {des
  Cloizeaux}}\ and\ \bibinfo {author} {\bibfnamefont {J.~J.}\ \bibnamefont
  {Pearson}},\ }\bibfield  {title} {\bibinfo {title} {{Spin-Wave Spectrum of
  the Antiferromagnetic Linear Chain}},\ }\href
  {https://doi.org/10.1103/PhysRev.128.2131} {\bibfield  {journal} {\bibinfo
  {journal} {Phys. Rev.}\ }\textbf {\bibinfo {volume} {128}},\ \bibinfo {pages}
  {2131} (\bibinfo {year} {1962})}\BibitemShut {NoStop}%
\bibitem [{\citenamefont {Thielemann}\ \emph {et~al.}(2009)\citenamefont
  {Thielemann}, \citenamefont {R\"uegg}, \citenamefont {R\o{}nnow},
  \citenamefont {L\"auchli}, \citenamefont {Caux}, \citenamefont {Normand},
  \citenamefont {Biner}, \citenamefont {Kr\"amer}, \citenamefont {G\"udel},
  \citenamefont {Stahn}, \citenamefont {Habicht}, \citenamefont {Kiefer},
  \citenamefont {Boehm}, \citenamefont {McMorrow},\ and\ \citenamefont
  {Mesot}}]{thielemann2009a}%
  \BibitemOpen
  \bibfield  {author} {\bibinfo {author} {\bibfnamefont {B.}~\bibnamefont
  {Thielemann}}, \bibinfo {author} {\bibfnamefont {C.}~\bibnamefont {R\"uegg}},
  \bibinfo {author} {\bibfnamefont {H.~M.}\ \bibnamefont {R\o{}nnow}}, \bibinfo
  {author} {\bibfnamefont {A.~M.}\ \bibnamefont {L\"auchli}}, \bibinfo {author}
  {\bibfnamefont {J.-S.}\ \bibnamefont {Caux}}, \bibinfo {author}
  {\bibfnamefont {B.}~\bibnamefont {Normand}}, \bibinfo {author} {\bibfnamefont
  {D.}~\bibnamefont {Biner}}, \bibinfo {author} {\bibfnamefont {K.~W.}\
  \bibnamefont {Kr\"amer}}, \bibinfo {author} {\bibfnamefont {H.-U.}\
  \bibnamefont {G\"udel}}, \bibinfo {author} {\bibfnamefont {J.}~\bibnamefont
  {Stahn}}, \bibinfo {author} {\bibfnamefont {K.}~\bibnamefont {Habicht}},
  \bibinfo {author} {\bibfnamefont {K.}~\bibnamefont {Kiefer}}, \bibinfo
  {author} {\bibfnamefont {M.}~\bibnamefont {Boehm}}, \bibinfo {author}
  {\bibfnamefont {D.~F.}\ \bibnamefont {McMorrow}},\ and\ \bibinfo {author}
  {\bibfnamefont {J.}~\bibnamefont {Mesot}},\ }\bibfield  {title} {\bibinfo
  {title} {Direct {Observation} of {Magnon} {Fractionalization} in the
  {Quantum} {Spin} {Ladder}},\ }\href
  {https://doi.org/10.1103/PhysRevLett.102.107204} {\bibfield  {journal}
  {\bibinfo  {journal} {Phys. Rev. Lett.}\ }\textbf {\bibinfo {volume} {102}},\
  \bibinfo {pages} {107204} (\bibinfo {year} {2009})}\BibitemShut {NoStop}%
\bibitem [{\citenamefont {Lake}\ \emph {et~al.}(2013)\citenamefont {Lake},
  \citenamefont {Tennant}, \citenamefont {Caux}, \citenamefont {Barthel},
  \citenamefont {Schollw\"ock}, \citenamefont {Nagler},\ and\ \citenamefont
  {Frost}}]{lake2013}%
  \BibitemOpen
  \bibfield  {author} {\bibinfo {author} {\bibfnamefont {B.}~\bibnamefont
  {Lake}}, \bibinfo {author} {\bibfnamefont {D.~A.}\ \bibnamefont {Tennant}},
  \bibinfo {author} {\bibfnamefont {J.-S.}\ \bibnamefont {Caux}}, \bibinfo
  {author} {\bibfnamefont {T.}~\bibnamefont {Barthel}}, \bibinfo {author}
  {\bibfnamefont {U.}~\bibnamefont {Schollw\"ock}}, \bibinfo {author}
  {\bibfnamefont {S.~E.}\ \bibnamefont {Nagler}},\ and\ \bibinfo {author}
  {\bibfnamefont {C.~D.}\ \bibnamefont {Frost}},\ }\bibfield  {title} {\bibinfo
  {title} {{Multispinon Continua at Zero and Finite Temperature in a Near-Ideal
  Heisenberg Chain}},\ }\href {https://doi.org/10.1103/PhysRevLett.111.137205}
  {\bibfield  {journal} {\bibinfo  {journal} {Phys. Rev. Lett.}\ }\textbf
  {\bibinfo {volume} {111}},\ \bibinfo {pages} {137205} (\bibinfo {year}
  {2013})}\BibitemShut {NoStop}%
\bibitem [{\citenamefont {Klanj\v{s}ek}\ \emph {et~al.}(2008)\citenamefont
  {Klanj\v{s}ek}, \citenamefont {Mayaffre}, \citenamefont {Berthier},
  \citenamefont {Horvati\'c}, \citenamefont {Chiari}, \citenamefont
  {Piovesana}, \citenamefont {Bouillot}, \citenamefont {Kollath}, \citenamefont
  {Orignac}, \citenamefont {Citro},\ and\ \citenamefont
  {Giamarchi}}]{klanjsek2008}%
  \BibitemOpen
  \bibfield  {author} {\bibinfo {author} {\bibfnamefont {M.}~\bibnamefont
  {Klanj\v{s}ek}}, \bibinfo {author} {\bibfnamefont {H.}~\bibnamefont
  {Mayaffre}}, \bibinfo {author} {\bibfnamefont {C.}~\bibnamefont {Berthier}},
  \bibinfo {author} {\bibfnamefont {M.}~\bibnamefont {Horvati\'c}}, \bibinfo
  {author} {\bibfnamefont {B.}~\bibnamefont {Chiari}}, \bibinfo {author}
  {\bibfnamefont {O.}~\bibnamefont {Piovesana}}, \bibinfo {author}
  {\bibfnamefont {P.}~\bibnamefont {Bouillot}}, \bibinfo {author}
  {\bibfnamefont {C.}~\bibnamefont {Kollath}}, \bibinfo {author} {\bibfnamefont
  {E.}~\bibnamefont {Orignac}}, \bibinfo {author} {\bibfnamefont
  {R.}~\bibnamefont {Citro}},\ and\ \bibinfo {author} {\bibfnamefont
  {T.}~\bibnamefont {Giamarchi}},\ }\bibfield  {title} {\bibinfo {title}
  {Controlling {Luttinger} {Liquid} {Physics} in {Spin} {Ladders} under a
  {Magnetic} {Field}},\ }\href {https://doi.org/10.1103/PhysRevLett.101.137207}
  {\bibfield  {journal} {\bibinfo  {journal} {Phys. Rev. Lett.}\ }\textbf
  {\bibinfo {volume} {101}},\ \bibinfo {pages} {137207} (\bibinfo {year}
  {2008})}\BibitemShut {NoStop}%
\bibitem [{\citenamefont {Kestin}\ and\ \citenamefont
  {Giamarchi}(2019)}]{kestin2019}%
  \BibitemOpen
  \bibfield  {author} {\bibinfo {author} {\bibfnamefont {N.}~\bibnamefont
  {Kestin}}\ and\ \bibinfo {author} {\bibfnamefont {T.}~\bibnamefont
  {Giamarchi}},\ }\bibfield  {title} {\bibinfo {title} {Low-dimensional
  correlations under thermal fluctuations},\ }\href
  {https://doi.org/10.1103/PhysRevB.99.195121} {\bibfield  {journal} {\bibinfo
  {journal} {Phys. Rev. B}\ }\textbf {\bibinfo {volume} {99}},\ \bibinfo
  {pages} {195121} (\bibinfo {year} {2019})}\BibitemShut {NoStop}%
\bibitem [{\citenamefont {Ewings}\ \emph {et~al.}(2016)\citenamefont {Ewings},
  \citenamefont {Buts}, \citenamefont {Le}, \citenamefont {Duijn},
  \citenamefont {Bustinduy},\ and\ \citenamefont {Perring}}]{ewings2016}%
  \BibitemOpen
  \bibfield  {author} {\bibinfo {author} {\bibfnamefont {R.~A.}\ \bibnamefont
  {Ewings}}, \bibinfo {author} {\bibfnamefont {A.}~\bibnamefont {Buts}},
  \bibinfo {author} {\bibfnamefont {M.~D.}\ \bibnamefont {Le}}, \bibinfo
  {author} {\bibfnamefont {J.~v.}\ \bibnamefont {Duijn}}, \bibinfo {author}
  {\bibfnamefont {I.}~\bibnamefont {Bustinduy}},\ and\ \bibinfo {author}
  {\bibfnamefont {T.~G.}\ \bibnamefont {Perring}},\ }\bibfield  {title}
  {\bibinfo {title} {Horace: {Software} for the analysis of data from single
  crystal spectroscopy experiments at time-of-flight neutron instruments},\
  }\href {https://doi.org/http://dx.doi.org/10.1016/j.nima.2016.07.036}
  {\bibfield  {journal} {\bibinfo  {journal} {Nucl. Instrum. Methods Phys.
  Res., Sect. A}\ }\textbf {\bibinfo {volume} {834}},\ \bibinfo {pages} {132}
  (\bibinfo {year} {2016})}\BibitemShut {NoStop}%
\bibitem [{\citenamefont {Schollw\"ock}(2011)}]{schollwock2011}%
  \BibitemOpen
  \bibfield  {author} {\bibinfo {author} {\bibfnamefont {U.}~\bibnamefont
  {Schollw\"ock}},\ }\bibfield  {title} {\bibinfo {title} {The density-matrix
  renormalization group in the age of matrix product states},\ }\href
  {https://doi.org/https://doi.org/10.1016/j.aop.2010.09.012} {\bibfield
  {journal} {\bibinfo  {journal} {Ann. Phys.}\ }\textbf {\bibinfo {volume}
  {326}},\ \bibinfo {pages} {96} (\bibinfo {year} {2011})}\BibitemShut
  {NoStop}%
\bibitem [{\citenamefont {Kim}\ \emph {et~al.}(1996)\citenamefont {Kim},
  \citenamefont {Matsuura}, \citenamefont {Shen}, \citenamefont {Motoyama},
  \citenamefont {Eisaki}, \citenamefont {Uchida}, \citenamefont {Tohyama},\
  and\ \citenamefont {Maekawa}}]{kim1996}%
  \BibitemOpen
  \bibfield  {author} {\bibinfo {author} {\bibfnamefont {C.}~\bibnamefont
  {Kim}}, \bibinfo {author} {\bibfnamefont {A.~Y.}\ \bibnamefont {Matsuura}},
  \bibinfo {author} {\bibfnamefont {Z.-X.}\ \bibnamefont {Shen}}, \bibinfo
  {author} {\bibfnamefont {N.}~\bibnamefont {Motoyama}}, \bibinfo {author}
  {\bibfnamefont {H.}~\bibnamefont {Eisaki}}, \bibinfo {author} {\bibfnamefont
  {S.}~\bibnamefont {Uchida}}, \bibinfo {author} {\bibfnamefont
  {T.}~\bibnamefont {Tohyama}},\ and\ \bibinfo {author} {\bibfnamefont
  {S.}~\bibnamefont {Maekawa}},\ }\bibfield  {title} {\bibinfo {title}
  {{Observation of Spin-Charge Separation in One-Dimensional SrCuO$_2$}},\
  }\href {https://doi.org/10.1103/PhysRevLett.77.4054} {\bibfield  {journal}
  {\bibinfo  {journal} {Phys. Rev. Lett.}\ }\textbf {\bibinfo {volume} {77}},\
  \bibinfo {pages} {4054} (\bibinfo {year} {1996})}\BibitemShut {NoStop}%
\bibitem [{\citenamefont {Kim}\ \emph {et~al.}(1997)\citenamefont {Kim},
  \citenamefont {Shen}, \citenamefont {Motoyama}, \citenamefont {Eisaki},
  \citenamefont {Uchida}, \citenamefont {Tohyama},\ and\ \citenamefont
  {Maekawa}}]{kim1997}%
  \BibitemOpen
  \bibfield  {author} {\bibinfo {author} {\bibfnamefont {C.}~\bibnamefont
  {Kim}}, \bibinfo {author} {\bibfnamefont {Z.-X.}\ \bibnamefont {Shen}},
  \bibinfo {author} {\bibfnamefont {N.}~\bibnamefont {Motoyama}}, \bibinfo
  {author} {\bibfnamefont {H.}~\bibnamefont {Eisaki}}, \bibinfo {author}
  {\bibfnamefont {S.}~\bibnamefont {Uchida}}, \bibinfo {author} {\bibfnamefont
  {T.}~\bibnamefont {Tohyama}},\ and\ \bibinfo {author} {\bibfnamefont
  {S.}~\bibnamefont {Maekawa}},\ }\bibfield  {title} {\bibinfo {title}
  {Separation of spin and charge excitations in one-dimensional {SrCuO$_2$}},\
  }\href {https://doi.org/10.1103/PhysRevB.56.15589} {\bibfield  {journal}
  {\bibinfo  {journal} {Phys. Rev. B}\ }\textbf {\bibinfo {volume} {56}},\
  \bibinfo {pages} {15589} (\bibinfo {year} {1997})}\BibitemShut {NoStop}%
\bibitem [{\citenamefont {Kim}\ \emph {et~al.}(2006)\citenamefont {Kim},
  \citenamefont {Koh}, \citenamefont {Rotenberg}, \citenamefont {Oh},
  \citenamefont {Eisaki}, \citenamefont {Motoyama}, \citenamefont {Uchida},
  \citenamefont {Tohyama}, \citenamefont {Maekawa}, \citenamefont {Shen},\ and\
  \citenamefont {Kim}}]{kim2006}%
  \BibitemOpen
  \bibfield  {author} {\bibinfo {author} {\bibfnamefont {B.~J.}\ \bibnamefont
  {Kim}}, \bibinfo {author} {\bibfnamefont {H.}~\bibnamefont {Koh}}, \bibinfo
  {author} {\bibfnamefont {E.}~\bibnamefont {Rotenberg}}, \bibinfo {author}
  {\bibfnamefont {S.-J.}\ \bibnamefont {Oh}}, \bibinfo {author} {\bibfnamefont
  {H.}~\bibnamefont {Eisaki}}, \bibinfo {author} {\bibfnamefont
  {N.}~\bibnamefont {Motoyama}}, \bibinfo {author} {\bibfnamefont
  {S.}~\bibnamefont {Uchida}}, \bibinfo {author} {\bibfnamefont
  {T.}~\bibnamefont {Tohyama}}, \bibinfo {author} {\bibfnamefont
  {S.}~\bibnamefont {Maekawa}}, \bibinfo {author} {\bibfnamefont {Z.-X.}\
  \bibnamefont {Shen}},\ and\ \bibinfo {author} {\bibfnamefont
  {C.}~\bibnamefont {Kim}},\ }\bibfield  {title} {\bibinfo {title} {Distinct
  spinon and holon dispersions in photoemission spectral functions from
  one-dimensional {SrCuO$_2$}},\ }\href {https://doi.org/10.1038/nphys316}
  {\bibfield  {journal} {\bibinfo  {journal} {Nature Phys.}\ }\textbf {\bibinfo
  {volume} {2}},\ \bibinfo {pages} {397} (\bibinfo {year} {2006})}\BibitemShut
  {NoStop}%
\bibitem [{\citenamefont {Nayak}\ \emph {et~al.}(2020)\citenamefont {Nayak},
  \citenamefont {Blosser}, \citenamefont {Zheludev},\ and\ \citenamefont
  {Mila}}]{nayak2020}%
  \BibitemOpen
  \bibfield  {author} {\bibinfo {author} {\bibfnamefont {M.}~\bibnamefont
  {Nayak}}, \bibinfo {author} {\bibfnamefont {D.}~\bibnamefont {Blosser}},
  \bibinfo {author} {\bibfnamefont {A.}~\bibnamefont {Zheludev}},\ and\
  \bibinfo {author} {\bibfnamefont {F.}~\bibnamefont {Mila}},\ }\bibfield
  {title} {\bibinfo {title} {{Magnetic-Field-Induced Bound States in
  Spin-$\frac12$ Ladders}},\ }\href
  {https://doi.org/10.1103/PhysRevLett.124.087203} {\bibfield  {journal}
  {\bibinfo  {journal} {Phys. Rev. Lett.}\ }\textbf {\bibinfo {volume} {124}},\
  \bibinfo {pages} {087203} (\bibinfo {year} {2020})}\BibitemShut {NoStop}%
\bibitem [{\citenamefont {Sorella}\ and\ \citenamefont
  {Parola}(1998)}]{sorella1998}%
  \BibitemOpen
  \bibfield  {author} {\bibinfo {author} {\bibfnamefont {S.}~\bibnamefont
  {Sorella}}\ and\ \bibinfo {author} {\bibfnamefont {A.}~\bibnamefont
  {Parola}},\ }\bibfield  {title} {\bibinfo {title} {Theory of hole propagation
  in one-dimensional insulators and superconductors},\ }\href
  {https://doi.org/10.1103/PhysRevB.57.6444} {\bibfield  {journal} {\bibinfo
  {journal} {Phys. Rev. B}\ }\textbf {\bibinfo {volume} {57}},\ \bibinfo
  {pages} {6444} (\bibinfo {year} {1998})}\BibitemShut {NoStop}%
\bibitem [{\citenamefont {Saiga}\ and\ \citenamefont
  {Kuramoto}(1999)}]{saiga1999}%
  \BibitemOpen
  \bibfield  {author} {\bibinfo {author} {\bibfnamefont {Y.}~\bibnamefont
  {Saiga}}\ and\ \bibinfo {author} {\bibfnamefont {Y.}~\bibnamefont
  {Kuramoto}},\ }\bibfield  {title} {\bibinfo {title} {{Dynamical Properties of
  the One-Dimensional Supersymmetric $t$-$J$ Model: A View from Elementary
  Excitations}},\ }\href {https://doi.org/10.1143/JPSJ.68.3631} {\bibfield
  {journal} {\bibinfo  {journal} {J. Phys. Soc. Jpn.}\ }\textbf {\bibinfo
  {volume} {68}},\ \bibinfo {pages} {3631} (\bibinfo {year}
  {1999})}\BibitemShut {NoStop}%
\bibitem [{\citenamefont {White}\ and\ \citenamefont
  {Affleck}(2008)}]{white2008}%
  \BibitemOpen
  \bibfield  {author} {\bibinfo {author} {\bibfnamefont {S.~R.}\ \bibnamefont
  {White}}\ and\ \bibinfo {author} {\bibfnamefont {I.}~\bibnamefont
  {Affleck}},\ }\bibfield  {title} {\bibinfo {title} {{Spectral function for
  the $S = 1$ Heisenberg antiferromagetic chain}},\ }\href
  {https://doi.org/10.1103/PhysRevB.77.134437} {\bibfield  {journal} {\bibinfo
  {journal} {Phys. Rev. B}\ }\textbf {\bibinfo {volume} {77}},\ \bibinfo
  {pages} {134437} (\bibinfo {year} {2008})}\BibitemShut {NoStop}%
\bibitem [{\citenamefont {Pereira}\ \emph {et~al.}(2009)\citenamefont
  {Pereira}, \citenamefont {White},\ and\ \citenamefont
  {Affleck}}]{pereira2009}%
  \BibitemOpen
  \bibfield  {author} {\bibinfo {author} {\bibfnamefont {R.~G.}\ \bibnamefont
  {Pereira}}, \bibinfo {author} {\bibfnamefont {S.~R.}\ \bibnamefont {White}},\
  and\ \bibinfo {author} {\bibfnamefont {I.}~\bibnamefont {Affleck}},\
  }\bibfield  {title} {\bibinfo {title} {{Spectral function of spinless
  fermions on a one-dimensional lattice}},\ }\href
  {https://doi.org/10.1103/PhysRevB.79.165113} {\bibfield  {journal} {\bibinfo
  {journal} {Phys. Rev. B}\ }\textbf {\bibinfo {volume} {79}},\ \bibinfo
  {pages} {165113} (\bibinfo {year} {2009})}\BibitemShut {NoStop}%
\bibitem [{\citenamefont {Barthel}\ \emph {et~al.}(2009)\citenamefont
  {Barthel}, \citenamefont {Schollw\"ock},\ and\ \citenamefont
  {White}}]{barthel2009}%
  \BibitemOpen
  \bibfield  {author} {\bibinfo {author} {\bibfnamefont {T.}~\bibnamefont
  {Barthel}}, \bibinfo {author} {\bibfnamefont {U.}~\bibnamefont
  {Schollw\"ock}},\ and\ \bibinfo {author} {\bibfnamefont {S.~R.}\ \bibnamefont
  {White}},\ }\bibfield  {title} {\bibinfo {title} {Spectral functions in
  one-dimensional quantum systems at finite temperature using the density
  matrix renormalization group},\ }\href
  {https://doi.org/10.1103/PhysRevB.79.245101} {\bibfield  {journal} {\bibinfo
  {journal} {Phys. Rev. B}\ }\textbf {\bibinfo {volume} {79}},\ \bibinfo
  {pages} {245101} (\bibinfo {year} {2009})}\BibitemShut {NoStop}%
\bibitem [{\citenamefont {Matsueda}\ \emph {et~al.}(2005)\citenamefont
  {Matsueda}, \citenamefont {Bulut}, \citenamefont {Tohyama},\ and\
  \citenamefont {Maekawa}}]{matsueda2005}%
  \BibitemOpen
  \bibfield  {author} {\bibinfo {author} {\bibfnamefont {H.}~\bibnamefont
  {Matsueda}}, \bibinfo {author} {\bibfnamefont {N.}~\bibnamefont {Bulut}},
  \bibinfo {author} {\bibfnamefont {T.}~\bibnamefont {Tohyama}},\ and\ \bibinfo
  {author} {\bibfnamefont {S.}~\bibnamefont {Maekawa}},\ }\bibfield  {title}
  {\bibinfo {title} {{Temperature dependence of spinon and holon excitations in
  one-dimensional Mott insulators}},\ }\href
  {https://doi.org/10.1103/PhysRevB.72.075136} {\bibfield  {journal} {\bibinfo
  {journal} {Phys. Rev. B}\ }\textbf {\bibinfo {volume} {72}},\ \bibinfo
  {pages} {075136} (\bibinfo {year} {2005})}\BibitemShut {NoStop}%
\bibitem [{\citenamefont {Kollath}\ \emph {et~al.}(2005)\citenamefont
  {Kollath}, \citenamefont {Schollw\"ock},\ and\ \citenamefont
  {Zwerger}}]{kollath2005}%
  \BibitemOpen
  \bibfield  {author} {\bibinfo {author} {\bibfnamefont {C.}~\bibnamefont
  {Kollath}}, \bibinfo {author} {\bibfnamefont {U.}~\bibnamefont
  {Schollw\"ock}},\ and\ \bibinfo {author} {\bibfnamefont {W.}~\bibnamefont
  {Zwerger}},\ }\bibfield  {title} {\bibinfo {title} {{Spin-Charge Separation
  in Cold Fermi Gases: A Real Time Analysis}},\ }\href
  {https://doi.org/10.1103/PhysRevLett.95.176401} {\bibfield  {journal}
  {\bibinfo  {journal} {Phys. Rev. Lett}\ }\textbf {\bibinfo {volume} {95}},\
  \bibinfo {pages} {176401} (\bibinfo {year} {2005})}\BibitemShut {NoStop}%
\bibitem [{\citenamefont {Feiguin}\ and\ \citenamefont
  {Fiete}(2010)}]{feiguin2010}%
  \BibitemOpen
  \bibfield  {author} {\bibinfo {author} {\bibfnamefont {A.~E.}\ \bibnamefont
  {Feiguin}}\ and\ \bibinfo {author} {\bibfnamefont {G.}~\bibnamefont
  {Fiete}},\ }\bibfield  {title} {\bibinfo {title} {{Spectral properties of a
  spin-incoherent Luttinger liquid}},\ }\href
  {https://doi.org/10.1103/PhysRevB.81.075108} {\bibfield  {journal} {\bibinfo
  {journal} {Phys. Rev. B}\ }\textbf {\bibinfo {volume} {81}},\ \bibinfo
  {pages} {075108} (\bibinfo {year} {2010})}\BibitemShut {NoStop}%
\bibitem [{\citenamefont {Cheianov}\ and\ \citenamefont
  {Zvonarev}(2004)}]{cheianov2004}%
  \BibitemOpen
  \bibfield  {author} {\bibinfo {author} {\bibfnamefont {V.~V.}\ \bibnamefont
  {Cheianov}}\ and\ \bibinfo {author} {\bibfnamefont {M.~B.}\ \bibnamefont
  {Zvonarev}},\ }\bibfield  {title} {\bibinfo {title} {{Nonunitary Spin-Charge
  Separation in a One-Dimensional Fermion Gas}},\ }\href
  {https://doi.org/10.1103/PhysRevLett.92.176401} {\bibfield  {journal}
  {\bibinfo  {journal} {Phys. Rev. Lett.}\ }\textbf {\bibinfo {volume} {92}},\
  \bibinfo {pages} {176401} (\bibinfo {year} {2004})}\BibitemShut {NoStop}%
\bibitem [{\citenamefont {Fiete}(2007)}]{fiete2007}%
  \BibitemOpen
  \bibfield  {author} {\bibinfo {author} {\bibfnamefont {G.}~\bibnamefont
  {Fiete}},\ }\bibfield  {title} {\bibinfo {title} {{The spin-incoherent
  Luttinger liquid}},\ }\href {https://doi.org/10.1103/RevModPhys.79.801}
  {\bibfield  {journal} {\bibinfo  {journal} {Rev. Mod. Phys.}\ }\textbf
  {\bibinfo {volume} {79}},\ \bibinfo {pages} {801} (\bibinfo {year}
  {2007})}\BibitemShut {NoStop}%
\end{thebibliography}%

\vspace{12pt}
\noindent
{\bf{Acknowledgements}}
\newline
\noindent
We thank T. Barthel, C. Berthod, F. Essler, V. Gorbenko, T. Guidi, N. A. Kamar,
D. McMorrow, M. Mena, M. Spira and S. Takayoshi for helpful discussions. 
This work was supported in part by the European Community Seventh
Framework Programme (FP7/2007-2013) under Grant Agreement 
No.~290605 (PSI-FELLOW/COFUND) (B.W.), by the Swiss National Science 
Foundation (SNF) under Grants No.~200020-132877 (Ch. R. and K.K.), 
No.~200020-188687 (T.G.) and No.~200020-219400 (T.G.) and by the Deutsche 
Forschungsgemeinschaft (DFG, German Research Foundation) under Grants
No.~277625399-TRR185 (B3) (C.K.), No.~511713970-CRC1639 NuMeriQS 
(C.K.) and No.~277146847-CRC1238 (C05) (C.K.), as well as through the Cluster 
of Excellence Matter and Light for Quantum Computing (ML4Q) EXC 
2004/1-390534769 (F.L.).
We acknowledge the provision of neutron beam-time at the ILL within proposal 
4-03-1598, at ISIS within experiment number RB1410142 and at the Swiss 
Spallation Neutron Source, SINQ, and we thank the technical and scientific staff 
for support at all three facilities. Calculations were performed at the University 
of Geneva on the Baobab and Mafalda clusters and at the University of Bonn on 
the BAF cluster.

\medskip
\noindent
{\bf {Data availability}}
\newline
\noindent
The raw INS datasets from this study are available at 
\href{https://doi.org/10.5286/ISIS.E.RB1410142}
{https://doi.org/10.5286/ISIS.E.RB1410142}. 
All l-MPS, c-MPS and processed INS data shown in the figures and otherwise supporting the 
results of the manuscript may be found at \href{https://doi.org/10.5281/zenodo.14989856}
{https://doi.org/10.5281/zenodo.14989856}. 

\medskip
\noindent
{\bf {Code availability}}
\newline
\noindent
Neutron scattering data were processed using the open-source software 
\textsc{Mantid}\citep{arnold2014} and \textsc{HORACE}\cite{ewings2016} 
as detailed in the Methods section. The codes used for MPS calculations 
are available from the corresponding author on request. 

\medskip
\noindent
{\bf {Author contributions}}
\newline
\noindent
B.W., Ch.R., A.L. and T.G. designed the study. D.B. and K.K. grew the single
crystals. B.W., S.W., B.T., M.B., R.B and Ch.R. performed the INS experiments 
and analysed the data. F.L., N.K., P.B. and B.W. performed MPS calculations 
with input from T.G., C.K., B.N. and A.L. The manuscript was written by B.W., 
F.L., C.K., A.L. and B.N. with input from all the authors.

\medskip
\noindent
{\bf {Additional information}}
\newline
\noindent
{\bf Competing interests:} the authors declare no competing interests.

\end{document}